\documentclass{aa}

\usepackage{graphicx}
\usepackage{txfonts}
\usepackage[separate-uncertainty = true]{siunitx}
\DeclareSIUnit{\dex}{dex}
\DeclareSIUnit{\Msun}{\textit{M}_\odot}
\DeclareSIUnit{\Rsun}{\textit{R}_\odot}
\DeclareSIUnit{\Lsun}{\textit{L}_\odot}
\DeclareSIUnit{\pc}{pc}
\DeclareSIUnit{\mas}{mas}
\DeclareSIUnit{\year}{yr}
\DeclareSIUnit{\mag}{mag}
\usepackage{lscape}
\usepackage{float}
\usepackage{natbib}
\usepackage[colorlinks,citecolor=blue]{hyperref}
\hypersetup{
  colorlinks = true,
  linkcolor = blue,
  anchorcolor = blue,
  citecolor = blue,
  filecolor = blue,
  urlcolor = blue}

\begin{document}
   \title{The present-day cosmic phosphorus abundance}

   \author{P. Aschenbrenner\inst{1}
          \and 
          K. Butler\inst{2}
          \and
          N. Przybilla\inst{1}
          }

   \institute{Institut f\"ur Astro- und Teilchenphysik, Universit\"at Innsbruck, Technikerstr. 25/8, 6020 Innsbruck, Austria\\
              \email{patrick.aschenbrenner@student.uibk.ac.at; norbert.przybilla@uibk.ac.at}
         \and
             LMU M\"unchen, Universit\"atssternwarte, Scheinerstr. 1, 81679 M\"unchen, Germany
             }

   \date{Received ; accepted }

  \abstract
   {}
   {The present-day phosphorus abundance in the solar neighbourhood is determined from a sample of OB-type stars. This is in order to constrain the endpoint of the galactochemical evolution of phosphorus in the course of stellar nucleosynthesis over cosmic time and to provide an abundance baseline for the study of the depletion of phosphorus onto dust grains in the interstellar medium.}
   {A model atom for \ion{P}{ii/iii/iv} based on a comprehensive new set of ab initio data for line transitions, photoionisations and electron-impact excitation was developed. Non-local thermodynamic equilibrium (non-LTE) line formation calculations with the codes {\sc Detail} and {\sc Surface} were conducted, based on LTE line-blanketed hydrostatic model atmospheres computed with the {\sc Atlas12} code. High-resolution optical spectra for a sample of 42 apparently slowly rotating main-sequence OB-type stars and B-type supergiants in the solar vicinity within $\sim$500\,pc and beyond, out to a distance of $\sim$2\,kpc, were analysed.}
   {The non-LTE effects on the formation of the \ion{P}{ii/iii/iv} lines are discussed. Non-LTE effects on the stellar abundances range from zero to $\sim$0.3\,dex. Where available in the spectra, ionisation balance between two phosphorus ionic species is achieved. Accurate and precise abundances are provided for the sample stars, statistical and systematic 1$\sigma$ uncertainties are typically each well below 0.1\,dex. The present-day cosmic phosphorus abundance in the solar neighbourhood  is constrained to $\log$\,(P/H)+12\,=\,5.36$\pm$0.14, which is compatible with the solar photospheric abundance, but lower than derived by LTE analyses of neutral phosphorus lines in solar-type stars. The amount of phosphorus depleted onto dust grains is $\sim$0.25\,dex.}
   {}

   \keywords{Atomic data -- Line: formation -- Stars: abundances -- Stars: atmospheres -- Stars: early-type -- Astrobiology}

   \maketitle

\section{Introduction}
As an element with an odd number of protons, phosphorus is one of the less abundant elements of the first three periods of the period system, appearing in the 18${th}$ position in the Solar System abundance distribution with $\log$\,(P/H)\,+\,12\,=\,5.43$\pm$0.03 \citep[meteoritic value,][]{Lodders21,Asplundetal21} and 11${th}$ in the Earth's crust \citep{Haynes16}. It is one of the essential elements for life on Earth and probably for astrobiology in general, together with hydrogen, carbon, nitrogen, oxygen, and sulphur. Phosphorus is one of the major constituents of cell membranes, of deoxyribonucleic acid (DNA) and ribonucleic acid (RNA), and is the main agent for energy exchange in cells based on adenosine tri- and diphosphate (ATP and ADP) (see e.g. \citealt{BuJe13}).
The single stable isotope of phosphorus, $^{31}$P, is produced during the late stages of the evolution of massive stars in the carbon- and neon-burning shells \citep{Clayton03} by the $^{30}$Si(p,$\gamma$)$^{31}$P reaction, later to be released in core-collapse supernovae \citep{Kooetal13}. Oxygen-neon (ONe) novae have also recently been suggested as a phosphorus source \citep{BeTs24}.

Phosphorus was detected in the gas-phase of the interstellar medium (ISM) by observation of ultraviolet \ion{P}{ii} lines decades ago \citep[e.g.][]{JuYo78,Jenkinsetal86}; see \citet{Jenkins09} for a re-evaluation of the data available in the literature. Some phosphorus-bearing molecules, eight species to date, have also been found in the ISM, in star-forming regions and in circumstellar material (see e.g. \citet{Fontani24} for an overview). The amount of interstellar phosphorus depleted onto dust is under debate. While \citet{JuYo78} suggested a depletion by a factor of two to three relative to the solar value in the diffuse ISM, \citet{Duftonetal86} found phosphorus undepleted in warm diffuse ISM clouds and depleted by a factor of three in cold diffuse clouds. Most recently, \citet{Ritcheyetal23} find phosphorus essentially undepleted in the diffuse ISM, whereas increasingly severe depletions are seen along molecule-rich sight lines. An important question in this context is related to the abundance reference for the evaluation of the depletion degree: the 4.6\,Gyr old Sun is potentially not the ideal abundance standard for the present-day ISM (see e.g. \citet{SnWi96}, \citet{Przybillaetal08}, and \citet{NiPr12}).

Data availability in the stellar case is quite different. The elemental abundances of phosphorus in long-lived solar-type stars required to study the galactochemical evolution of the element \citep{Cescuttietal12} have only been derived in recent years because of the absence of analysable spectral lines of \ion{P}{i} in the classical optical spectral range. However, spectral lines from two multiplets are present in the near-infrared $Y$ band \citep[e.g.][]{Caffauetal11,Caffauetal16,Maasetal17,Maasetal22,Snedenetal21,SaNi22} and one multiplet in the ultraviolet \citep{Jacobsonetal14,Roedereretal14}, which are accessible with the new generation of high-resolution infrared spectrographs and with the Hubble Space Telescope, respectively. Later, studies of \ion{P}{i} lines in the $H$ band have become available \citep{Hawkinsetal16,Caffauetal19,Nandakumaretal22}. To date, all of these analyses have been performed assuming that the line formation occurs under the conditions of local thermodynamic equilibrium (LTE). A first step to consider deviations from LTE (non-LTE effects) has been made only recently to derive the solar phosphorus abundance \citep[see their Appendix]{Takeda24}; however, he relies on a great deal of approximate atomic data due to the lack of more accurate ab initio data. The non-LTE abundance corrections were found to be small, about $-$0.04 to $-$0.02\,dex relative to the LTE values for the Sun. These investigations need to be extended on the basis of ab initio data and broadened towards stars with a larger effective temperature range and towards lower gravities and lower metallicity, which are more prone to non-LTE effects \citep[for some very first results see also][]{Takeda24}. 

Lines of \ion{P}{ii} and \ion{P}{iii} become visible in the optical spectra of hotter stars, most notably in the chemically peculiar mercury-manganese (HgMn) stars and the helium-weak phosphorus-gallium (PGa) stars,
where overabundance factors can reach a factor $\sim$100 times solar for phosphorus \citep[see e.g.][]{Smith96,GhAl16}. However, these overabundances are a consequence of atomic diffusion taking place in very stable atmospheres \citep{Michaud70}, rendering these stars useless for deriving pristine abundances.\footnote{Some phosphorus-rich stars are also known among cool low-mass giants \citep{Masseronetal20,Brauneretal23}, but the overabundance in that case is likely inherited from another nearby stellar source.}

More suitable are early B-type and late O-type stars, which have mass-loss rates sufficiently strong to avoid chemical segregation \citep[with the exception of the rare magnetic helium-strong stars, see e.g.][]{Smith96}, but weak enough to prevent the photospheric layers from being significantly affected by the hydrodynamic outflow. Being short-lived, these OB-type stars typically do not travel far from their places of birth (with the exception of runaway stars). At the same time, they are luminous, and are therefore excellent targets to trace pristine present-day elemental abundances throughout the Milky Way. 

Although phosphorus lines were identified almost one hundred years ago in the normal B2\,IV star $\gamma$~Pegasi \citep{Struve30}, very few studies on phosphorus abundances in normal OB-type stars are available in the literature. The sharp-lined B3\,IV star $\iota$~Herculis has several phosphorus determinations in the optical \citep{PeAl70,PePo85,PiAd93} and in the ultraviolet \citep{GoLa17}; $\gamma$ Pegasi has been similarly investigated \citep{PiAd93,NiPo06}. More recently, phosphorus abundances in a few early B-type stars in the $\lambda$~Orionis region have been reported by \citet{ElUn23}. In addition, phosphorus abundances in a few later B-type stars with superficially normal abundances are available \citep{Fossatietal09,Niemczuraetal09} and for the late B-type supergiant $\beta$~Orionis \citep{Przybillaetal2006}, which has evolved from a late O-type star on the main sequence. All analyses were performed assuming LTE and often the atmospheric parameter determination involved photometric techniques. Typically, the dispersion from the line-by-line analysis is large in the studies, and reveals a mismatch of abundances from different ionisation stages and a wide range of phosphorus abundances from sub- to super-solar in different stars and even in the same star is found, even though they are relatively young and are all located in the solar neighbourhood.

The first non-LTE abundance study of \ion{P}{ii} for a sample of normal B-type stars as well as chemically peculiar stars, in total 85 objects, has only recently been provided by \citet{Takeda24}. The phosphorus abundance from his sample of normal B-stars is near-solar in LTE, but sub-solar by $\sim$0.2 to 0.3\,dex in non-LTE, apparently contradicting the scenario of Galactic chemical evolution. As was the case for \ion{P}{i}, most atomic input data for the \ion{P}{ii} model atom relied on approximations, only the \ion{P}{ii} $\lambda$6043.084\,{\AA} line was analysed for each star and the atmospheric parameters of the sample stars were derived from Str\"omgren photometry. Overall, there is much room for improvement.

In the present work we therefore approach the problem from the ground up. A model atom for \ion{P}{ii/iii/iv} is constructed from a comprehensive set of new ab initio data, allowing all accessible phosphorus lines in high-quality optical spectra to be modelled, sometimes from two ionisation stages simultaneously. All sample stars have been thoroughly investigated previously, so that accurate and precise atmospheric parameters are available from quantitative spectroscopy. The aim of the paper is to derive the present-day cosmic abundance of phosphorus in the solar neighbourhood, a region that encompasses the local Orion Arm out to a distance of 500\,pc and that includes the nearest star-forming regions delineated by the Gould Belt. This provides the endpoint of the Galactic chemical evolution of phosphorus in the course of stellar nucleosynthesis over cosmic time, allows for a comparison with the Solar System value, and gives an abundance baseline for the study of the depletion of phosphorus onto dust grains in the interstellar medium. Knowledge of the cosmic phosphorus abundance is crucial for the field of astrobiology, the study of habitability of exoplanets, and the development of life \citep{Hinkeletal20}.

The paper is structured as follows. The construction of a new phosphorus model atom as the basis for all further model calculations is discussed in Sect.~\ref{sect:modelling}, and the non-LTE effects in Sect.~\ref{sect:non-LTE}. Details on the spectral analysis of the observational data are summarised in Sect.~\ref{sect:analysis}, and the results in Sect.~\ref{sect:results}. Our conclusions are drawn in Sect.~\ref{sect:summary}. Some assessment of the ab initio data is made in Appendix~\ref{appendix:A}, a summary of the parameters for each individual object analysed is found in Appendix~\ref{appendix:B}, and the results from the line-by-line analysis of the sample stars are tabulated in Appendix~\ref{appendix:C}.

\section{Model calculations}\label{sect:modelling}

\subsection{Model atmospheres and programmes}
The spectral analysis was based on a hybrid non-local thermodynamic equilibrium (non-LTE) approach. First, the atmospheric structure assuming LTE was computed with the {\sc Atlas9} and {\sc Atlas12} codes \citep{Kurucz93,Kurucz05}. Then, non-LTE line-formation calculations with updated and extended versions of the programmes {\sc Detail} and {\sc Surface} \citep{giddings81,BuGi85} were performed to calculate synthetic model spectra, which were compared to observations using self-written Python routines, aiming at $\chi^2$-mimimisation. This hybrid approach has previously been used to determine stellar parameters and chemical abundances of a variety of hot stars, including slowly rotating single OB main-sequence stars \citep[e.g.][]{NiPr07,NiPr12,Aschenbrenneretal23} and massive B-type supergiants \citep[e.g.][]{Wessmayeretal22}.

\subsection{The model atom}
\subsubsection{Energy levels} 
Our model atom for the trace species phosphorus considers the ionisation stages \ion{P}{ii/iii/iv} plus the ground state of \ion{P}{v}; \ion{P}{i} was omitted because of its negligible occupation in the atmospheres of OB stars. For \ion{P}{ii} all energy states up to 6$s$\,$^1$P\degr\ were explicitly considered, from the singlet and triplet spin systems and one quintet state. They were packed into 45 $LS$-coupled terms. For \ion{P}{iii} all states in the doublet and quartet spin systems up to 6$s$\,$^2$S were included, packed into 25 terms. For \ion{P}{iv}, all singlet and triplet states up to 5$s$\,$^1$S were considered, packed into 23 terms. All level energies were adopted from \citet{Martinetal85}. 

\begin{table}[t!]
\centering
\caption{Phosphorus lines present in the optical spectra of OB stars.}
\label{table:p_lines}
\small
\setlength{\tabcolsep}{1.2mm}
\begin{tabular}{lccrccr}
\hline\hline
Ion          & $\lambda$\,(\AA) & Mult. & $\chi$\,(eV) & lower level & upper level & $\log gf$ \\
\hline \\[-3mm]
\ion{P}{ii}\tablefootmark{a}  & 4475.270 & 24 & 13.09 & 3$p$4$p$\,$^3$P$_2$         & 3$p$4$d$\,$^3$D$^{\circ}_3$ & 0.428\\
\ion{P}{ii}  & 4499.230 & ...& 13.38 &  3$p$4$p$\,$^1$D$_2$        & 3$p$4$d$\,$^1$F$^{\circ}_3$ & 0.495 \\
\ion{P}{ii}  & 4969.701 & 13 & 12.81 &  3$p$4$p$\,$^3$D$_2$        & 3$p$5$s$\,$^3$P$^{\circ}_1$ & $-$0.119 \\
\ion{P}{ii}  & 5253.479 & 10 & 11.02 & 3$p$4$s$\,$^1$P$^{\circ}_1$ & 3$p$4$p$\,$^1$D$_2$         &    0.287 \\
\ion{P}{ii}  & 5296.077 &  7 & 10.80 & 3$p$4$s$\,$^3$P$^{\circ}_2$ & 3$p$4$p$\,$^3$S$_1$         & $-$0.124 \\
\ion{P}{ii}  & 5386.895 &  6 & 10.75 & 3$p$4$s$\,$^3$P$^{\circ}_1$ & 3$p$4$p$\,$^3$P$_1$         & $-$0.320 \\
\ion{P}{ii}  & 5425.880 &  6 & 10.80 & 3$p$4$s$\,$^3$P$^{\circ}_2$ & 3$p$4$p$\,$^3$P$_2$         &    0.223 \\
\ion{P}{ii}  & 6024.178 &  5 & 10.75 & 3$p$4$s$\,$^3$P$^{\circ}_1$ & 3$p$4$p$\,$^3$D$_2$         &    0.132 \\
\ion{P}{ii}\tablefootmark{b}  & 6043.084 &  5 & 10.80 & 3$p$4$s$\,$^3$P$^{\circ}_2$ & 3$p$4$p$\,$^3$D$_3$         &    0.375 \\[1mm]
\ion{P}{iii} & 4059.312 &  1 & 14.49 & 3$d$\,$^2$D$_{2.5}$         & 4$p$\,$^2$P$^{\circ}_{1.5}$ & $-$0.228 \\
\ion{P}{iii} & 4080.084 &  1 & 14.49 & 3$d$\,$^2$D$_{1.5}$         & 4$p$\,$^2$P$^{\circ}_{0.5}$ & $-$0.484 \\
\ion{P}{iii}\tablefootmark{a} & 4222.195 &  3 & 14.61 & 4$s$\,$^2$S$_{0.5}$         & 4$p$\,$^2$P$^{\circ}_{1.5}$ &    0.131 \\
\ion{P}{iii} & 4246.720 &  3 & 14.61 & 4$s$\,$^2$S$_{0.5}$         & 4$p$\,$^2$P$^{\circ}_{0.5}$ & $-$0.172 \\[1mm]
\ion{P}{iv}\tablefootmark{b}  & 3347.739 &  1 & 28.13 & 3$s$4$s$\,$^3$S$_1$         & 3$s$4$p$\,$^3$P$^{\circ}_2$ &    0.251 \\
\ion{P}{iv}  & 3364.470 &  1 & 28.13 & 3$s$4$s$\,$^3$S$_1$         & 3$s$4$p$\,$^3$P$^{\circ}_1$ &    0.022 \\
\ion{P}{iv}  & 3371.119 &  1 & 28.13 & 3$s$4$s$\,$^3$S$_1$         & 3$s$4$p$\,$^3$P$^{\circ}_0$ & $-$0.451 \\
\ion{P}{iv}\tablefootmark{b}  & 4249.655 &  2 & 29.01 & 3$s$4$s$\,$^1$S$_0$         & 3$s$4$p$\,$^1$P$^{\circ}_1$ & $-$0.130 \\[0.5mm]
\hline
\end{tabular}
\tablefoot{
\tablefoottext{a}{Strongly blended lines, omitted in the analysis.}
\tablefoottext{b}{Weakly blended lines, included in the analysis.}
}
\end{table}

\subsubsection{Radiative transitions}
The oscillator strengths and photoionisation cross-sections for all
radiatively permitted transitions between the energy terms considered
were computed by one of us (Butler, in prep.). The ab initio calculations employed
the close-coupling approximation of electron-atom collision theory
using the $R$-matrix method \citep{Berringtonetal87}, the same
techniques as used by the Opacity Project \citep{Seatonetal94}. 
An assessment of some aspects regarding the quality of the ab initio data is made in Appendix~\ref{appendix:A}.
At the
stage of the calculations with {\sc Detail} the 358, 81, and 61 bound-bound
transitions between terms of \ion{P}{ii}, \ion{P}{iii}, and \ion{P}{iv},
respectively were implemented using
Doppler profiles. The resonance structure of the photoionisation
cross-sections was sampled using a grid of over 60\,000
frequency points.

For the final spectrum synthesis with {\sc Surface}, fine-structure transition data \citep{FFTI06} were employed that originated from multiconfiguration Hartree-Fock (MCHF) calculations, which account for relativistic effects through the use of the Breit-Pauli Hamiltonian. Transition wavelengths were taken from \citet{Readeretal80} and coefficients for natural broadening and broadening by the quadratic Stark effect from Kurucz.\footnote{\url{http://kurucz.harvard.edu/atoms.html}}
Table~\ref{table:p_lines} summarises the data on the analysed lines: ion, transition wavelength $\lambda$, multiplet number \citep[][]{Moore72}, excitation potential of the lower level $\chi$, designations of the lower and upper level of the transition, and the adopted oscillator strength $\log gf$. Several of the lines are blended with spectral features from other species, which in most cases is \ion{Fe}{iii}. Depending on the $T_{\rm eff}$ sensitive strength of the blending line and the separation, some phosphorus lines are unusable for the quantitative analysis by these blends.

\subsubsection{Collisional transitions}
A large set of $R$-matrix electron-impact excitation data was calculated for this project by one of us (Butler, in prep.), tabulated as effective collision strengths $\Upsilon$ for a wide temperature range. This comprises 861, 276, and 252 transitions between the different terms in \ion{P}{ii}, \ion{P}{iii}, and \ion{P}{iv}, respectively.
For the few remaining collisional bound-bound transitions Van Regemorter’s formula \citep{vanRegemorter62} was applied to radiatively permitted transitions adopting the previously described oscillator strengths, and for the optically forbidden transitions the semi-empirical Allen formula \citep{Allen73} was used with the collision strength $\Omega$ set to 1.0. All collisional ionisation data were calculated using the Seaton formula \citep{Seaton62}, adopting threshold photoionisation cross-sections from the data described above.

\begin{table}[t!]
\centering
\caption{LTE and non-LTE phosphorus abundances for selected stars.}
\label{table:LTE_abundance}
\small
\setlength{\tabcolsep}{1.8mm}
\begin{tabular}{llcccc}
\hline\hline
Name & Sp. Type & $T_{\mathrm{eff}}$ & $\log g$ & \multicolumn{2}{c}{$\log$\,(P/H)\,+\,12} \\[.3mm] \cline{5-6}
\rule{0mm}{.3mm}\\[-3mm]
     &          & K                  & (cgs)    & LTE & non-LTE\\    
\hline \\[-3mm]
HD\,57682  & O9.2\,IV & 33500 & 3.93 & 5.55$\pm$0.08 & 5.30$\pm$0.07 \\
HD\,149438 & B0.2\,V  & 32000 & 4.30 & 5.65$\pm$0.09 & 5.38$\pm$0.08 \\
HD\,36959  & B1\,V    & 26100 & 4.25 & 5.17$\pm$0.02 & 5.24$\pm$0.03 \\
HD\,886    & B2\,IV   & 22000 & 3.95 & 5.31$\pm$0.05 & 5.25$\pm$0.02 \\
HD\,160762 & B3\,IV   & 17500 & 3.80 & 5.42$\pm$0.14 & 5.40$\pm$0.13 \\
HD\,34085  & B8\,Ia   & 12000 & 1.75 & 5.52$\pm$0.05 & 5.53$\pm$0.07 \\
\hline
\end{tabular}
\tablefoot{Parameter uncertainties and references are listed in Table \ref{table:results}.}
\end{table}

\section{Discussion of the non-LTE effects}\label{sect:non-LTE}
We selected six stars from the whole sample, among which are the standards $\tau$~Sco (HD\,149438), $\gamma$~Peg (HD\,886), $\iota$~Her (HD\,160762), and $\beta$~Ori (Rigel, HD\,34085). The sample stars cover a wide range of effective temperature values and surface gravities. We performed for those six stars an abundance analysis in LTE in addition to non-LTE in order to investigate the difference between the two approaches. Table~\ref{table:LTE_abundance} lists the names of the stars, their spectral types, effective temperatures $T_\mathrm{eff}$, surface gravities $\log g$, LTE abundances, and non-LTE abundances. In hot stars, where \ion{P}{iv} lines are used for the analysis, the LTE calculation produces significantly weaker lines, implying up to $\sim$\,0.3\,dex higher phosphorus abundances in LTE. In cooler stars we find only small deviations ($\lesssim$\,0.05\,dex) from LTE.

\begin{figure}
    \centering
    \includegraphics[width=\linewidth]{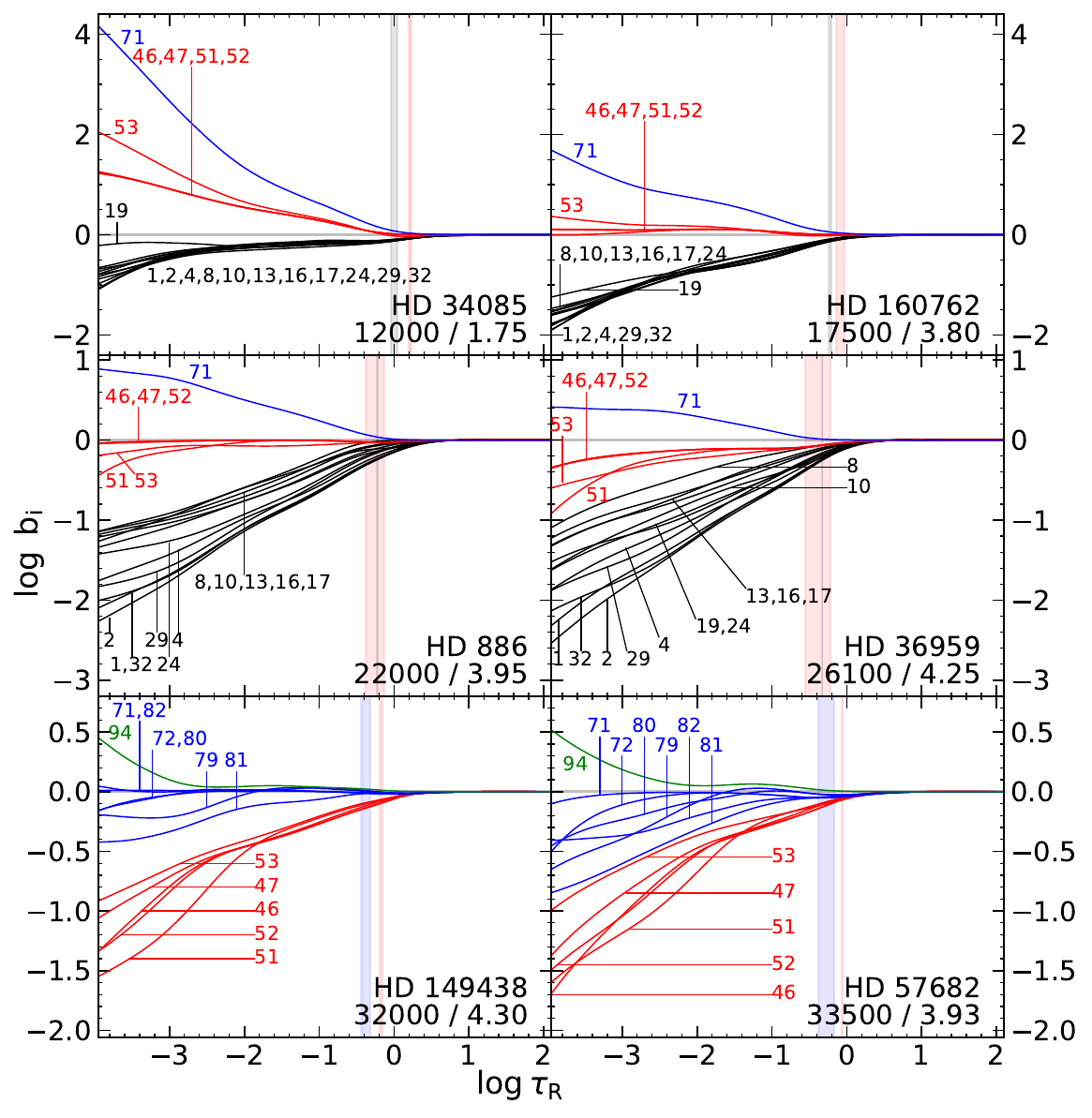}
    \caption{Departure coefficients $b_i$ for \ion{P}{ii} (black), \ion{P}{iii} (red), \ion{P}{iv} (blue), and \ion{P}{v} (green) levels as a function of the Rosseland optical depth $\tau_{\mathrm{R}}$ for the six selected sample stars. In the bottom right of each plot the name of the star together with its $T_\mathrm{eff}$ and $\log g$-value are given. The colour-matching shaded areas indicate the formation depth of the strongest observed line for each ion. The curves are labelled with the identifiers listed in Table~\ref{table:terms}.}
    \label{fig:dep_coef}
\end{figure}

The departure coefficients $b_i$\,=\,$n_i^{\rm NLTE}/n_i^{\rm LTE}$, defined as the ratio of level populations $n_i$ in non-LTE to LTE for energy levels $i$, are shown in Fig.~\ref{fig:dep_coef} as a function of the Rosseland optical depth $\tau_\mathrm{R}$. We show all energy levels involved for the formation of the observed spectral lines, as well as the ground state energy level of the next higher ion for which no lines are present (see Table~\ref{table:terms} for identifiers of the $LS$-coupled terms).
Deep in the stellar atmosphere, where the density increases and collisional processes dominate, the departure coefficients approach unity, that is, they fulfil the inner LTE boundary condition. The levels of the major ionisation stage \ion{P}{ii} are underpopulated in the outer atmosphere of HD\,34085 because of photoionisations, leading to a non-LTE overpopulation of \ion{P}{iii} and to a much higher overpopulation of the \ion{P}{iv} ground state (though the absolute occupation remains low for this minor ionisation species). The \ion{P}{ii} levels become further overionised as $T_\mathrm{eff}$ increases, and this behaviour repeats itself throughout the different higher ionisation stages to the late O-type stars. The main ionisation stage remains closest to LTE and the higher minor ionisation stage is overpopulated. The lines are formed in the deepest visible 
\begin{table}[th]
\centering
\caption{$LS$-coupled term identifiers.}
\label{table:terms}
\small
\setlength{\tabcolsep}{1.2mm}
\begin{tabular}{lrllrllrl}
\hline\hline
Ion & No. & Term & Ion & No. & Term & Ion & No. & Term \\
\hline\\[-3mm]
\ion{P}{ii} &  1 & 3$p^2$\,$^3$P           & \ion{P}{ii}  & 19 & 4$p$\,$^1$D           & \ion{P}{iii} & 53 & 4$p$\,$^2$P$^{\circ}$     \\
            &  2 & 3$p^2$\,$^1$D           &              & 24 & 5$s$\,$^3$P$^{\circ}$ & \ion{P}{iv}  & 71 & 3$s^2$\,$^1$S             \\
            &  4 & 3$p^3$\,$^5$S$^{\circ}$ &              & 29 & 4$d$\,$^3$D$^{\circ}$ &              & 72 & 3$s$3$p$\,$^3$P$^{\circ}$ \\
            &  8 & 4$s$\,$^3$P$^{\circ}$   &              & 32 & 4$d$\,$^1$F$^{\circ}$ &              & 79 & 3$s$4$s$\,$^3$S           \\
            & 10 & 4$s$\,$^1$P$^{\circ}$   & \ion{P}{iii} & 46 & 3$p$\,$^2$P$^{\circ}$ &              & 80 & 3$s$4$s$\,$^1$S           \\
            & 13 & 4$p$\,$^3$D             &              & 47 & 3$s$3$p^2$\,$^4$P     &              & 81 & 3$s$4$p$\,$^3$P$^{\circ}$ \\
            & 16 & 4$p$\,$^3$P             &              & 51 & 3$d$\,$^2$D           &              & 82 & 3$s$4$p$\,$^1$P$^{\circ}$ \\
            & 17 & 4$p$\,$^3$S             &              & 52 & 4$s$\,$^2$S           & \ion{P}{v}   & 94 & 3$s$\,$^2$S               \\
\hline
\end{tabular}
\end{table}
photospheric regions where the departure coefficients deviate only slightly from LTE. We note that some of the spectral lines for which the formation regions are shown are not employed for the analysis because they are too weak at the signal-to-noise ratio ($S/N$)  of the observed spectra.

\begin{figure}
    \centering
    \includegraphics[width=\linewidth]{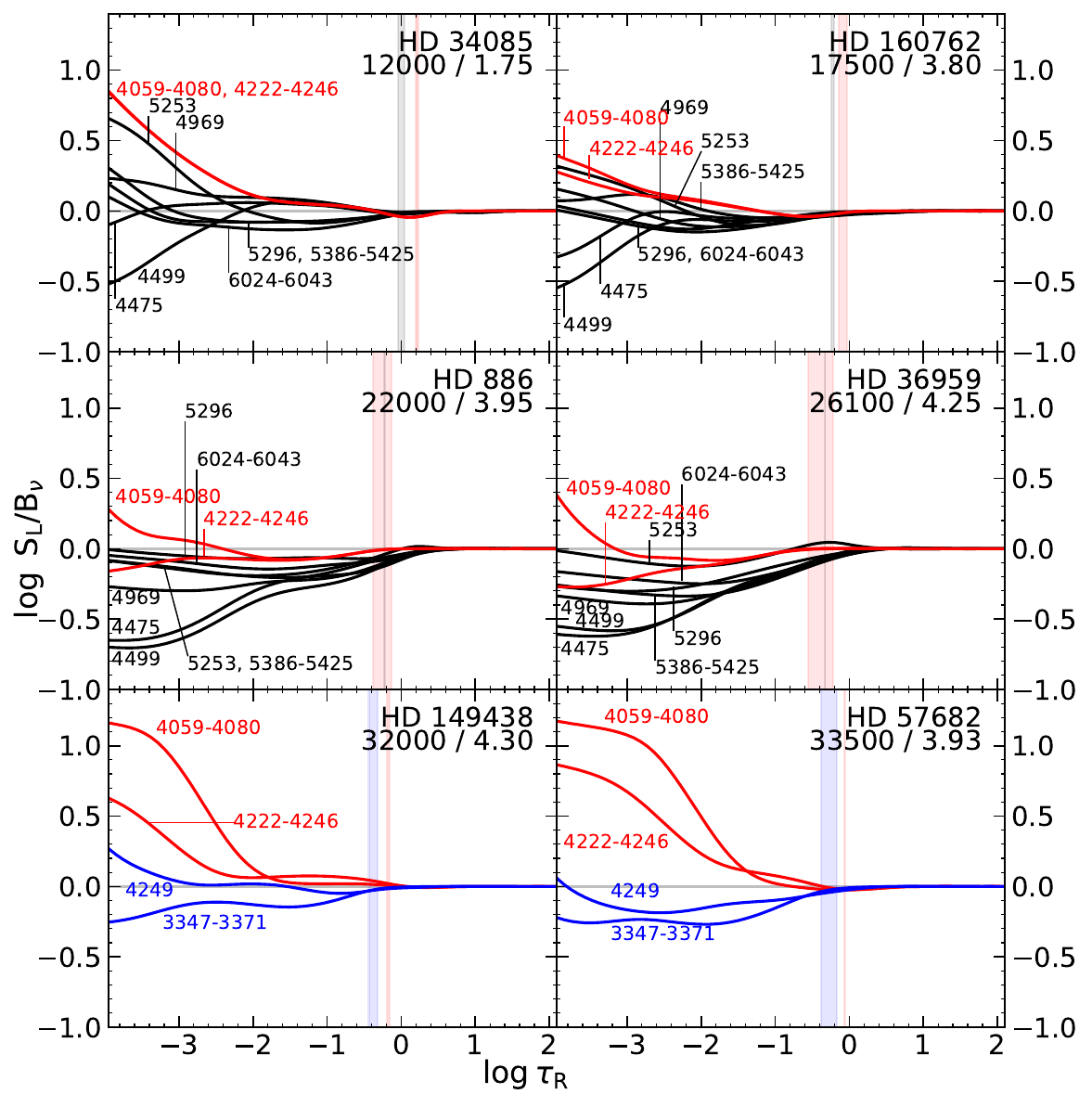}
    \caption{Ratio of the line source function $S_{\mathrm{L}}$ to the Planck function $B_{\nu}$ for \ion{P}{ii} (black), \ion{P}{iii} (red), and \ion{P}{iv} (blue) lines as a function of $\tau_{\mathrm{R}}$ for selected sample stars. The colour-matching shaded areas indicate the formation depth of the strongest observed line for each ion.}
    \label{fig:source_function}
\end{figure}

Figure~\ref{fig:source_function} shows the ratio of the line source function to the Planck function
\begin{align}
    \frac{S_{\rm L}}{B_{\nu}} = \frac{\exp(h\nu_{lu}/kT)-1}{(n_l^{\rm NLTE}g_u)/(n_u^{\rm NLTE}g_l)-1},
\end{align}
where $h$ denotes the Planck constant, $\nu_{lu}$ is the transition frequency, $k$ the Boltzmann constant, $T$ the temperature, $n_l$ and $n_u$ are respectively the population of the lower and upper energy levels of the transition, and $g_l$ and $g_u$ are the corresponding statistical weights. As the formation of the weak P lines occurs close to the continuum-forming regions, deviations of $S_\mathrm{L}/B_\nu$ from unity are small. Typically, the lines appear slightly deeper in non-LTE, that is, $S_\mathrm{L}/B_\nu$\,$\lesssim$\,1, leading to slightly lower abundance values derived in the analysis compared to LTE values. As shown in the two lower panels in Fig.~\ref{fig:source_function}, the effect is strongest for the \ion{P}{iv} lines.

\section{Spectral analysis}\label{sect:analysis}
\subsection{Observational data}
The spectral analysis is based on high-resolution spectra with a resolving power of $R$\,=\,$\Delta\lambda/\lambda$\,$\approx$\,40\,000 to 70\,000 and a high $S/N$, typically of several hundred, observed with different instruments.
For 15 stars we used spectra observed with the Fibre Optics Cassegrain Echelle Spectrograph \citep[{FOCES},][]{Pfeifferetal98} on the 2.2\,m telescope at the Calar Alto Observatory in Spain; 
14 stars were observed with the Fiberfed Extended Range Optical Spectrograph \citep[{FEROS},][]{Kauferetal99} on the Max-Planck-Gesellschaft/European Southern Observatory (ESO) 2.2\,m telescope at La Silla in Chile.

Nine spectra were observed with the Fiber-fed Echelle Spectrograph \citep[{FIES},][]{Telting2014} on the 2.5\,m Nordic Optical Telescope at La Palma;
two stars were observed with the Echelle Spectro-Polari\-metric Device for the Observation of Stars \citep[{ESPaDOnS},][]{ManDon03} on the 3.6\,m Canada-France-Hawaii telescope (CFHT) at Mauna Kea/Hawaii;
and one spectrum was taken with the Calar Alto Fiber-fed Echelle Spectrograph \citep[{CAFE},][]{Aceituno2013} on the 2.2\,m telescope at the Calar Alto Observatory in Spain.
The data reduction of most spectra is described in previous works (see the references in Table~\ref{table:results} for details). The spectra cover the optical wavelength range where most of the phosphorus lines are located (see Table~\ref{table:p_lines}). An exception is a triplet of \ion{P}{iv} lines, which lie in the optical-UV. For six stars, data in the optical-UV
were found in the ESO archive, obtained with the Ultraviolet and Visual Echelle Spectrograph \citep[{UVES},][]{Dekker2000} on the ESO Very Large Telescope UT-2 at Paranal in Chile. We downloaded the pipeline-reduced Phase~3 data from the ESO Science Portal\footnote{\url{https://archive.eso.org/scienceportal/home}} and normalised the spectral regions around the phosphorus lines with a linear continuum. 

We also investigated two additional samples of B-type supergiants, which we had investigated previously for the presence of phosphorus lines. The early B-type supergiants discussed by \citet{Wessmayeretal23,Wessmayeretal24} show too much broadening for the phosphorus lines to be visible. On the other hand, the late B-type stars discussed by \citet{FiPr12} are too cool to excite the \ion{P}{ii} lines to a level where they become distinguishable from the continuum at the $S/N$ reached.

\subsection{Atmospheric parameters}
We had already analysed all of the stars in our sample and we adopted the atmospheric parameters from the references provided in Table~\ref{table:results}. Multiple ionisation equilibria and the Stark-broadened hydrogen lines were used to determine $T_\mathrm{eff}$ and $\log g$. The microturbulent velocity $\xi$ was constrained by requiring that the elemental abundance of individual lines in the spectral fitting are independent of the strengths of the lines. The (projected) rotational velocity $\varv \sin i$ and the (radial-tangential) macroturbulent velocity $\zeta$ were derived simultaneously by fitting the profiles of numerous metal lines. Some of the stars were originally analysed on the basis of lower-resolution spectra
and with different atomic data. For four stars where the originally found solutions show significant deviations from the spectra used in this work (comparing hydrogen, helium, and silicon lines), we refined the atmospheric parameters.

\begin{figure}
    \centering
    \includegraphics[width=0.99\linewidth]{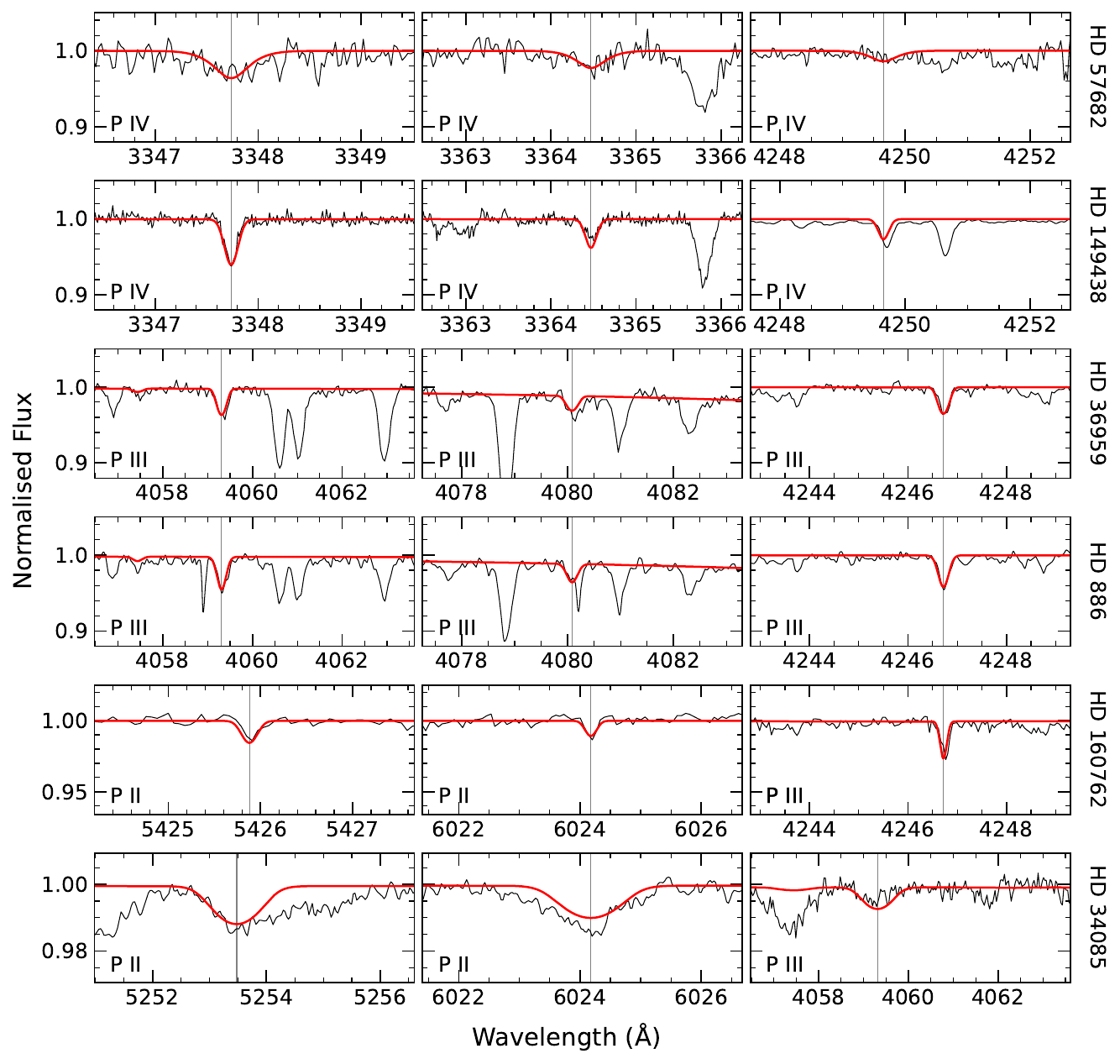}
    \caption{Global best fitting model shown for six different sample stars along the $T_\mathrm{eff}$-sequence (top to bottom). Each row shows the phosphorus lines for a single star. The line centres are marked by a grey vertical line. The ionic species of each line shown is shown at the bottom left of each panel.}
    \label{fig:fits}
\end{figure}

\begin{table*}[ht]
\centering
\caption{Estimation of systematic uncertainties in the non-LTE abundance analysis.}
\label{table:uncertainties}
\small
\begin{tabular}{lccccccccccc}
\hline\hline
 & \multicolumn{2}{c}{15000/2.60} & \multicolumn{2}{c}{15000/4.00} & \multicolumn{2}{c}{20000/4.00} & \multicolumn{1}{c}{25000/4.00} & \multicolumn{2}{c}{30000/4.00} & \multicolumn{1}{c}{34000/4.00} & \\
 & \ion{P}{ii} & \ion{P}{iii} & \ion{P}{ii} & \ion{P}{iii} & \ion{P}{ii} & \ion{P}{iii} & \ion{P}{iii} & \ion{P}{iii} & \ion{P}{iv} & \ion{P}{iv} & \\
\hline
$T_{\mathrm{eff}}$ $\pm$ $\SI{300}{\kelvin}$ & $\pm$0.05 & $\mp$0.07 & $\pm$0.01 & $\mp$0.07 & $\pm$0.02 & $\mp$0.04 & $\pm$0.01 & $\pm$0.05 & $\mp$0.05 & $\pm$0.01 \\
$\log g$ $\pm$ $\SI{0.1}{\dex}$              & $\mp$0.02 & $\pm$0.06 &  $\pm$0.02 & $\pm$0.05 & $\pm$0.00 & $\pm$0.06 & $\pm$0.01 & $\mp$0.04 & $\pm$0.10 & $\pm$0.03 \\
$\xi$ $\pm$ $\SI{1}{\kilo\meter\per\second}$ & $\mp$0.01 & $\mp$0.01 & $\mp$0.01 & $\mp$0.01 & $\mp$0.01 & $\mp$0.02 & $\mp$0.01 & $\pm$0.02 & $\mp$0.02 & $\mp$0.01 \\
$[$Fe/H$]$ $\pm$ $\SI{0.1}{\dex}$            & $+$0.02 & $\mp$0.02 & $+$0.01 & $\mp$0.01 & $+$0.01 & $\mp$0.01 & $+$0.02 & $+$0.02 & $\mp$0.02 & $+$0.02 \\
Continuum placement                          & $\pm$0.05 & $\pm$0.05 & $\pm$0.05 & $\pm$0.05 & $\pm$0.05 & $\pm$0.05 & $\pm$0.05 & $\pm$0.05 & $\pm$0.05 & $\pm$0.05 \\
\hline
Estimated total uncertainty                  & $\pm$0.07 & $\pm$0.08 & $\pm$0.06 & $\pm$0.09 & $\pm$0.06 & $\pm$0.08 & $\pm$0.06 & $\pm$0.08 & $\pm$0.11 & $\pm$0.07 \\
\hline
\end{tabular}
\end{table*}

\begin{figure}
    \includegraphics[width=\linewidth]{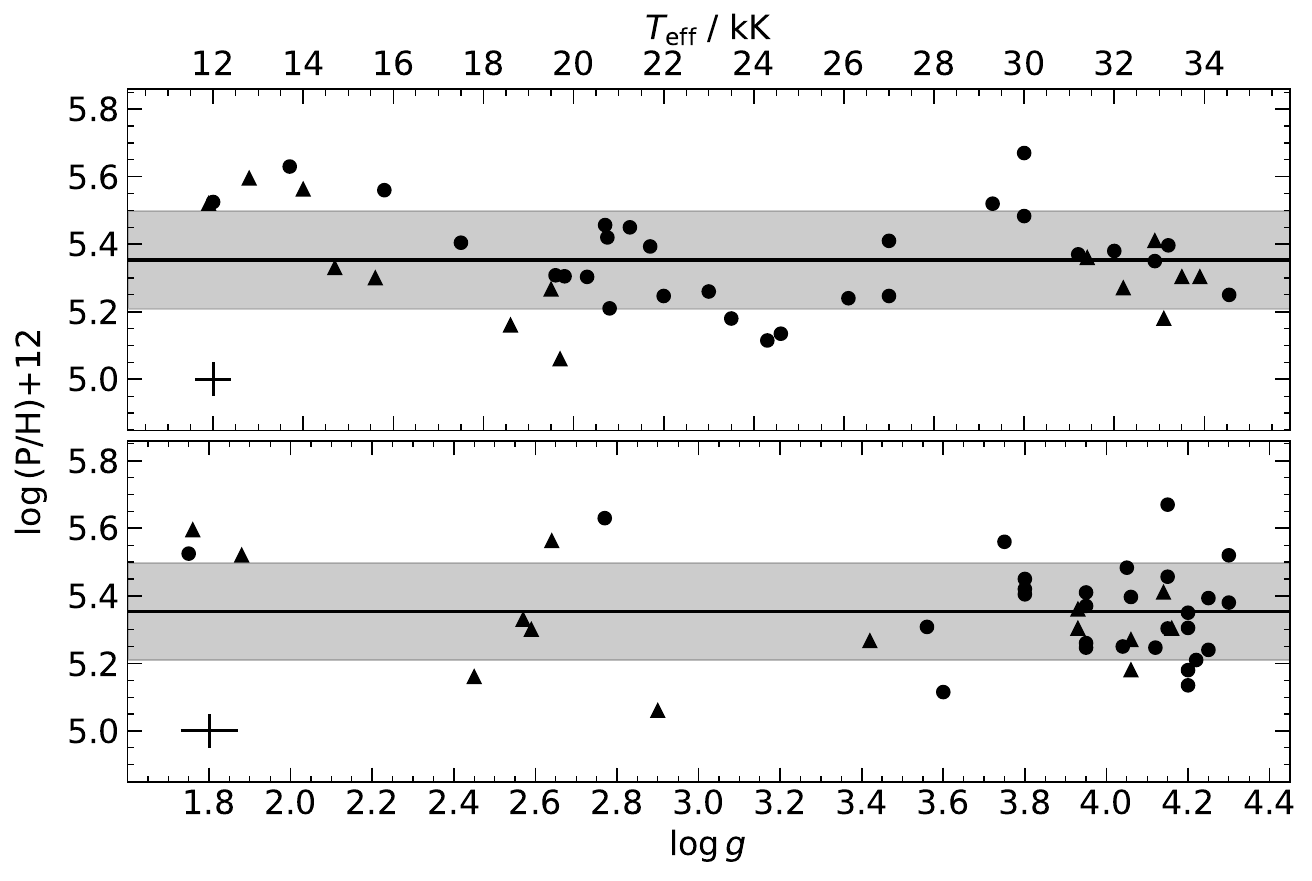}
    \caption{Phosphorus abundances as a function of fundamental stellar parameters for all sample stars. Upper panel: Phosphorus abundance as a function of $T_{\rm eff}$. Lower panel: Phosphorus abundance as a function of $\log g$. The black line shows the mean value, the shaded area indicates the 1$\sigma$ standard deviation. The typical uncertainties are indicated in the bottom left corner. Objects farther away than $\SI{500}{\pc}$ are marked with triangles; those closer are marked with dots.}
    \label{fig:teff_logg}
\end{figure}

\subsection{Abundance analysis}
The phosphorus abundance of each star is determined by fitting each spectral line individually.
A grid of model spectra for different phosphorus abundances ($\Delta \log$\,(P/H)\,+\,12\,=\,0.1\,dex) is calculated,
then the models are interpolated with a cubic spline to find the best fitting abundance value using $\chi^2$-minimisation for each line.
Finally, the mean and standard deviation of the individual line-by-line abundances are used as the stellar phosphorus abundance (see Table~\ref{table:results}).
Table \ref{table:line_fits} lists the line-by-line abundances for each star. In Fig.~\ref{fig:fits} we show examples for the resulting line fits for six different stars covering the atmospheric parameter range investigated here. A comparison between the observed spectrum and the model spectrum for the derived mean phosphorus abundance per star is shown. Overall, a good match is achieved simultaneously in all stars, with ionisation balances (typically \ion{P}{ii/iii}) being matched, where available.
We note, however, that the \ion{P}{iv} 4249.655 line appears to be redshifted in all spectra by about 0.05\,{\AA}, which can also be seen in Fig.~\ref{fig:fits}. The shift may either be an artefact from a blend with another line; however, we could not identify a spectral line at this wavelength from other ions present. More likely, the shift may be due to an experimental inaccuracy. \citet{ZeMa77} measured the wavelength to 4249.656\,{\AA},
but older work by \citet{Bowen32}, stating original measurements by \citet{Geuter07}, gives 4249.73\,{\AA}. In either case, to fit the abundance of this line, we corrected the model wavelength to match the observation.
From the example fits one can also see that phosphorus abundance determinations in OB-type stars are feasible only for slowly rotating objects, with a combined broadening from rotation and macroturbulence $\lesssim$50\,km\,s$^{-1}$. The weak phosphorus lines become indistinguishable from the continuum even at moderate rotational velocities.

\subsection{Systematic uncertainties}
In order to determine the effects of uncertainties in atmospheric parameters on the derived non-LTE phosphorus abundances, we performed test calculations for different combinations of $T_{\mathrm{eff}}$ and $\log g$ as listed in Table \ref{table:uncertainties}. All models were calculated with solar metallicity and $\xi$\,=\,$\SI{4}{\kilo\meter\per\second}$, except the model with $\log g$\,=\,2.60, where we used $\xi$\,=\,$\SI{10}{\kilo\meter\per\second}$. The effects of each individual parameter were determined by changing its value by a typical error margin. Then, with the new atmospheric parameter, a grid of models with different phosphorus abundances was calculated, and these models were used to fit the original spectrum. The difference in phosphorus abundance gave the estimated uncertainty. This investigation shows that a thorough determination of $T_\mathrm{eff}$ and $\log g$ is required if an accurate abundance determination is intended as the sensitivities to parameter variations are largest. The sensitivity to variations in microturbulence is small, as expected for weak lines. In addition, the effects of typical variations of the metallicity [Fe/H] are small.
An additional source of systematic error is the placement of the continuum. It depends on the $S/N$ and the strength of the spectral line considered. Our estimation of $\pm 0.05$\,dex is for spectra with $S/N$\,$\sim$\,100 \citep{PrBu01}, as typically achieved in the literature. We note that our $S/N$ values are much higher, so the placement of the continuum is less of a problem here.
The final total uncertainty was obtained via a Monte Carlo calculation. For each pair of $T_{\mathrm{eff}}$/$\log g$, atmospheric parameters were sampled from Gaussian distributions. For $\log T_{\mathrm{eff}}$
and $\log g$ we use a correlation of $\rho$\,=\,0.8, found by Markov chain Monte Carlo fitting of stellar spectra. The other parameters were treated as uncorrelated. Based on the difference between the new and original parameters, the uncertainties in Table~\ref{table:uncertainties} were linearly scaled and summed
to obtain the total uncertainty.

 Finally, we mention here that systematic uncertainties on the phosphorus abundances may also stem from uncertainties in the numerous atomic data used in the model atom (see Appendix~\ref{appendix:A} for further discussion).  Guided by previous investigations \citep[e.g.][]{Przybillaetal00,Przybillaetal01b,PrBu01} we can conclude that once high-quality atomic data are used for all relevant processes in the model atom construction, as is the case here, the impact of atomic data uncertainties on systematic uncertainties in abundance determinations is negligible if an average over several lines from different multiplets is available. The only direct effect stems from the uncertainties of the oscillator strengths used for the analysis of the observed spectral lines. The data in our case are from the MCHF calculations of \citet{FFTI06}, which should be accurate to a few per cent. Consequently, these also not affect the total estimated uncertainties (Table~\ref{table:uncertainties}). In summary, in the present work, the systematic effects on the determination of phosphorus abundance are less than 0.1\,dex.

\section{Results}\label{sect:results}
The results of our present investigations on slowly rotating chemically normal stars are summarised in Appendix~\ref{appendix:B}.
As a test for the presence of residual systematic uncertainties, we show in Fig.~\ref{fig:teff_logg} the abundances of phosphorus as a function of $T_\mathrm{eff}$ and $\log g$, respectively. The derived abundances show no correlation with either over the wide range of atmospheric parameters covered by our sample stars. As our present objects are distributed over larger distances than in the star sample used previously to establish the cosmic abundance standard (CAS) \citep{NiPr12,Przybillaetal08,Przybillaetal13}, objects closer than 500\,pc (i.e. in the solar neighbourhood) and beyond are marked by different symbols. This is because the presence of Galactic abundance gradients leads to higher abundances than CAS values at smaller galactocentric radii $R_\mathrm{G}$ than the solar $R_0$\,=\,8178$\pm$13(stat.)$\pm$22(sys.)\,pc \citep{GRAVITY19} and to lower abundances at larger $R_\mathrm{G}$.  However, the number of objects within and beyond $R_0$ is approximately equal and extends to less than about $\pm$2\,kpc (we therefore do not determine the phosphorus abundance gradient here), so that only an increase in the scatter around the mean value is to be expected. A mean value of $\log$\,(P/H)\,+\,12\,=\,5.35$\pm$0.14 is found for the entire sample of 42 stars. The scatter is larger than for the CAS values for other elements, which usually show a standard deviation of $\sim$0.05\,dex.\\[-8mm]
\paragraph{Present-day cosmic phosphorus abundance.}
 Taking into account only the 28 stars in the solar neighbourhood ($d$\,$\leq$\,500\,pc), a present-day cosmic phosphorus abundance can be established: $\log$\,(P/H)\,+\,12\,=\,5.36$\pm$0.14. It is interesting that the standard deviation is not reduced with respect to the full star sample implying that the larger abundance scatter is significant. The abundance of phosphorus is lower by more than an order of magnitude than any of those species that have been investigated to establish the CAS so far. Consequently, individual nucleosynthesis events that enrich the ISM may have a greater effect on the local abundance of the element than on those of the more abundant chemical species.

\begin{table}
\caption{Cosmic phosphorus abundances. }
\label{table:cosmic_abundances}
\setlength{\tabcolsep}{1.5mm}
\small
\centering
\begin{tabular}{lll}
\hline\hline
Object(s) & $\log$\,(P/H)\,+\,12 & Reference\\
\hline
Solar photosphere             & 5.41$\pm0.03$   & \citet{Asplundetal21}\\
CI meteorites                 & 5.43$\pm$0.03   & \citet{Lodders21}\\
Solar System (proto-solar)    & 5.52$\pm$0.03   & \citet{Lodders21}\\
solar-type stars             & 5.67$\pm$0.14  & \citet{Caffauetal19}\\
ISM gas ($d$\,$\leq$\,2\,kpc) & 5.12$\pm$0.09 & \citet{Ritcheyetal23}\\
B-ype stars                   & 5.20$\pm$0.18   & \citet{Takeda24}\\
OB stars                      & 5.36$\pm$0.14   & this work, CAS\\
\hline
\end{tabular}
\end{table}

The cosmic phosphorus abundances of other indicators are summarised in Table~\ref{table:cosmic_abundances}. Our CAS value is in good agreement with the solar photospheric and meteoritic abundances. Phosphorus abundances from solar-type stars in the solar neighbourhood seem to be systematically somewhat higher, but we note that these results originate from LTE analyses so that some downward revision from non-LTE effects may be possible \citep[though the effects appear to be small,][]{Takeda24}. Phosphorus abundances in the gas phase of the diffuse ISM are lower, as are data from mostly mid- to late B-type stars with superficially normal abundances. The different entries in Table~\ref{table:cosmic_abundances} will be discussed in various contexts in the following.\\[-8mm]
\begin{table}[t!]
\centering
\caption{Comparison of the present work with \citet{Takeda24}.}
\label{table:takeda}
\small
\setlength{\tabcolsep}{1.8mm}
\begin{tabular}{llcccc}
\hline\hline
Name & Sp. Type & $T_{\mathrm{eff}}$ & $\log g$ & \multicolumn{2}{c}{$\log$\,(P/H)\,+\,12}\\[.3mm] \cline{5-6}
\rule{0mm}{.3mm}\\[-3mm]
     &          & K                  & (cgs)    & LTE & non-LTE \\
\hline \\[-3mm]
HD\,886    & B2\,IV   & 22000 & 3.95 & 5.31$\pm$0.05 & 5.25$\pm$0.02 \\
           &          & 21667 & 3.83 & 5.32          & 5.28\\[.5mm]
HD\,35708  & B2.5\,IV & 20700 & 4.15 & ...           & 5.46$\pm$0.01 \\
           &          & 21082 & 4.09 & 5.68          & 5.43\\[.5mm]
HD\,35039  & B2\,V    & 19600 & 3.56 & ...           & 5.31$\pm$0.06 \\
           &          & 20059 & 3.69 & 5.34          & 5.20\\[.5mm]
HD\,160762 & B3\,IV   & 17500 & 3.80 & 5.42$\pm$0.14 & 5.40$\pm$0.13 \\
           &          & 17440 & 3.91 & 5.33          & 5.03\\[.5mm]
HD\,209008 & B3\,III  & 15800 & 3.75 & ...           & 5.56$\pm$0.05\\
           &          & 15353 & 3.50 & 5.53          & 5.17\\ 
\hline
\end{tabular}
\tablefoot{First line for each star: this work; second line: \citet{Takeda24}.}
\end{table}
\paragraph{Comparison with the literature data.}
The only other non-LTE study of phosphorus abundances in sharp-lined B-type stars is that of \citet{Takeda24}. His sample considers about 60 superficially normal stars with spectra that are comparable in quality to the present sample. Only the \ion{P}{ii} $\lambda$6043.084\,{\AA} line was analysed (a value of $\log gf$\,=\,+0.442 was adopted, which is larger than the value used in the present work by about 0.07\,dex), with the model atom therefore concentrating on \ion{P}{ii}. Apart from oscillator strengths taken from Kurucz and photoionisation cross-sections for the four energetically lowest terms, the atomic data in Takeda's model atom is approximate. The atmospheric parameters $T_\mathrm{eff}$ and $\log g$ of the sample stars were determined from Str\"omgren colours and should be accurate to 3\% in $T_\mathrm{eff}$ and 0.2\,dex in $\log g$. The microturbulent velocity was fixed at a constant value of $\xi$\,=\,1\,km\,s$^{-1}$ for $T_\mathrm{eff}$\,$<$16\,500\,K,~and~2\,km\,s$^{-1}$~above.

Takeda finds a mean phosphorus abundance for his sample slightly above solar in LTE, with a small systematic trend yielding higher values towards lower temperatures. The derived non-LTE abundance corrections are large, amounting to $-$0.1 to $-$0.5\,dex, leading to homogeneous abundances throughout his sample; however, they are appreciably underabundant relative to the Sun by $\sim$0.2 to 0.3\,dex. While there is agreement of the mean abundance with our value within the mutual (large) uncertainties, the systematic difference should be investigated. 

Five stars in our sample are in common with his so we can compare the results (see summary in Table~\ref{table:takeda}). There is agreement between the atmospheric parameters within the mutual uncertainties; the differences in $T_\mathrm{eff}$ and $\log g$ are in both directions, higher and lower. On the other hand, a systematic effect is seen in the derived abundances. While in our work only small non-LTE effects are present in the temperature range covered in Table~\ref{table:takeda} and good agreement of our abundances with the LTE results of Takeda is found, this does not apply to Takeda's non-LTE abundances. For HD\,886 there is still agreement between our abundance and Takeda's value, where Takeda analysed one extremely weak \ion{P}{ii} line with an equivalent width $W_\lambda$ of about 1\,m{\AA}, while we used the much stronger \ion{P}{iii} lines.
However, towards the cooler stars, where the measured $W_\lambda$ of the \ion{P}{ii} line reaches its maximum at about 10\,m{\AA} (which is still very weak), considerably larger non-LTE corrections are found in his work. Such large corrections are unexpected since the radiation field in these stars is not particularly intense and collisions deep in the atmospheres of main-sequence stars, where the lines are formed, should dominate. We cannot confirm the low phosphorus abundances of \citet{Takeda24}, and we can only speculate that his large non-LTE abundance corrections may be an artefact of predominantly approximate data used for the construction of his model atom.

\begin{figure}[ht]
    \centering
    \includegraphics[width=0.9\linewidth]{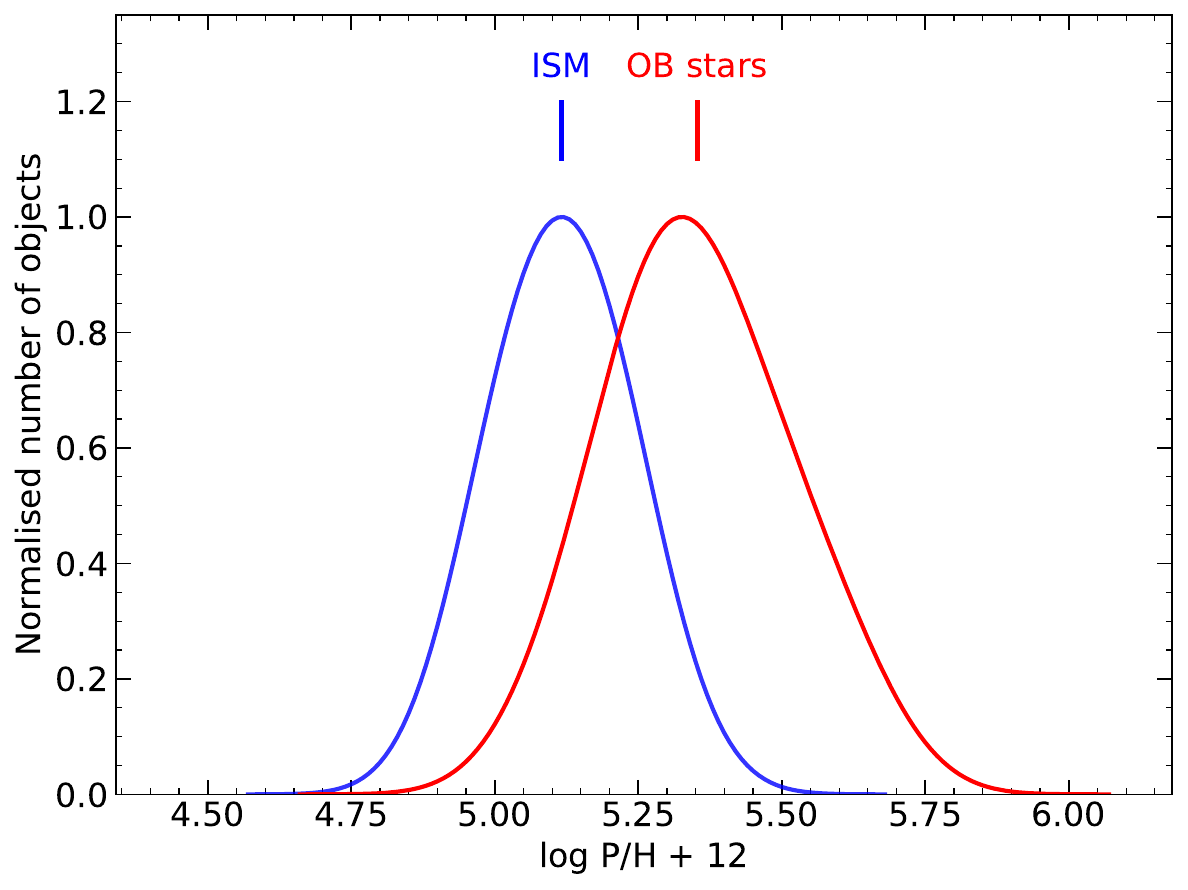}
    \caption{Kernel density estimate for the stellar phosphorus abundance distribution derived in this work (red) with
     gas-phase abundances along different sight lines ($d$\,$\leq$\,2\,kpc) of the diffuse ISM \citep[blue,][]{Ritcheyetal23}.
     The mean values are indicated. For both distributions a Gaussian kernel with $\sigma$\,=\,0.1 was used.}
    \label{fig:ism_dust}
\end{figure}

\paragraph{Depletion of phosphorus onto ISM dust grains.}
As discussed in the Introduction, the question regarding the amount of phosphorus depleted on dust grains in the diffuse ISM has so far not been settled conclusively. The phosphorus abundances derived here provide a solid basis to which ISM gas-phase abundances can be compared. For this comparison, we adopt a subsample of the data from \citet{Ritcheyetal23}. First, all of their sight lines with background stars at distances less than 2\,kpc were selected to match the region covered in the Milky Way and their column densities were transformed to the abundance scale employed here. This provided an asymmetric abundance distribution for the sample, with an extended tail towards low phosphorus abundances. Upon closer inspection of these sight lines, it became apparent that the background stars were main-sequence stars of spectral types B1 and later at distances of less than 700\,pc. For these sources, the ISM \ion{H}{i} column densities are low, while their stellar Ly$\alpha$ lines are comparatively broad, resulting in a higher deduced total hydrogen column density than is actually present. In consequence, the phosphorus abundances appear systematically too low. These sight lines are not discussed further. Two more sight lines with relatively nearby late O-type background stars showed similarly low phosphorus abundances and were also removed: \object{HD 24534} (X Per), a high-mass X-ray binary with a decretion disk around the fast-rotating visible star (which adds a complex Ly$\alpha$ signature) and \object{HD 73882}, which is the exciting star of the \ion{H}{ii} region \object{RCW 27}, and therefore--strictly speaking--does not trace the diffuse ISM alone. 

\begin{figure}[ht]
\centering
    \includegraphics[width=\linewidth]{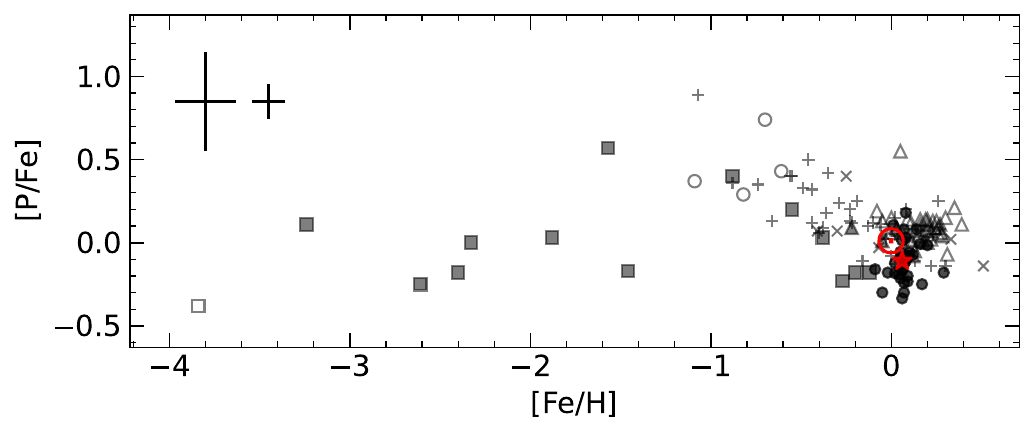}
    \caption{Cosmic evolution of [P/Fe], with [P/Fe]$<$1, as a function of [Fe/H]. Data from solar-type stars (\citet[filled and empty squares]{Roedereretal14, Roedereretal16}, \citet[filled and empty triangles]{Caffauetal16, Caffauetal19}, \citet[circles]{Masseronetal20}, \citet[crosses]{Maasetal22}, and \citet[plus signs]{Nandakumaretal22})  are compared to massive stars in the solar neighbourhood (our work, dots). The red five-pointed star marks the cosmic P abundance, the solar photospheric abundance \citep{Asplundetal21} is also indicated ($\odot$). Typical error bars are indicated in the top left corner. The larger crosses are for the five most metal-poor sources; the smaller crosses are for the others.}
    \label{fig:P_FE}
\end{figure}

The resulting kernel density estimation for the phosphorus abundance distributions in the diffuse ISM and in our star sample are compared in Fig.~\ref{fig:ism_dust}. The ISM abundance distribution is slightly tighter than the stellar distribution. However, in general, it implies that the abundances of phosphorus are less homogeneously distributed than the more abundant metals, which show tighter distributions \citep[see][]{NiPr12}. The difference between the two distributions of $\sim$0.25\,dex indicates the amount of phosphorus locked up in the dust grains; that is, approximately 40\% of the pristine phosphorus is incorporated into solids in the ISM.\\[-8mm]
\paragraph{Galactochemical evolution of phosphorus.}
Finally, we address our results in the context of the galactochemical evolution of phosphorus within the solar neighbourhood. The available observational data on the [P/Fe] ratios as a function of iron abundance ('metallicity'), with the bracket notation denoting normalisation to solar values, are displayed in Fig.~\ref{fig:P_FE}. Most of the data originate from long-lived solar-type stars, implying that [P/Fe] was slightly sub-solar in the early Milky Way; probably rose to values of about two to three times solar between [Fe/H] of $-2$ to $-1$; and since then has shown a drop, as is known from the $\alpha$-elements. Table~\ref{table:cosmic_abundances} shows that the average abundances of the solar-type stars are slightly super-solar; the spread in metallicity results from the migration of inner-disk metal-rich and (fewer) outer-disk metal-poor stars into the solar neighbourhood. Although the spread of phosphorus abundances in our sample is similar (given that the two distributions overlap) our stars show overall lower abundances. Much better agreement is seen if only the metal-rich specimens in the solar-type star samples of \citet[from UV spectra]{Roedereretal14} and of \citet[][]{Nandakumaretal22} are considered. We also note that our phosphorus abundances match the lower limit of data in the early Milky Way, where the element was produced in core-collapse supernovae of massive stars before nucleosynthesis in ONe novae led to a rise. Later still there was a downturn due to the reduction of the nova rate after the peak of star formation and by an increasing rate of supernovae of type Ia \citep{BeTs24}. However, further interpretation certainly requires thorough non-LTE analyses of the phosphorus abundances in solar-type stars to reduce potential systematic uncertainties. 

In conclusion, we briefly put the solar phosphorus abundance in the context of the CAS value. For this, one has to consider that photospheric abundances do not reflect the bulk abundances of the Sun because element settling takes place from the outer convection zone into the radiative zone of the Sun's envelope. Consequently, it is the proto-solar value of $\log$\,(P/H)\,+\,12\,=\,5.52$\pm$0.03 \citep{Lodders21} to which the mean abundance of phosphorus in our massive stars of $\log$\,(P/H)\,+\,12\,=\,5.36$\pm$0.14 must be compared, implying that the trend for a systematic difference is even greater. Because reliable data on the Galactic abundance gradient for phosphorus (our star sample is insufficient for a meaningful determination) and on its enrichment over the past 4.56\,Gyr (the age of the Sun) due to galactochemical evolution are not available, we cannot extend the discussion of \citet{NiPr12} for phosphorus on the birth place of the Sun in the Milky Way and its subsequent migration to its present location. However, we are confident that the difference between the proto-solar abundance and the CAS value for phosphorus will fit the picture of the birth place of the Sun being located at a galactocentric radius of 5 to 6\,kpc, as deduced from a number of other chemical elements.

\section{Summary and outlook}\label{sect:summary}
A model atom was developed for non-LTE line-formation calculations for \ion{P}{ii/iii/iv}, based on a comprehensive set of ab initio data. Non-LTE effects were found to be negligible in stars with effective temperatures below $\sim$20\,000\,K. At higher $T_\mathrm{eff}$ the non-LTE effects amount to $\sim$\,0.05\,dex for the \ion{P}{iii} lines, while the abundances of the \ion{P}{iv} lines change by up to $\sim$\,0.25\,dex. Phosphorus abundances for individual stars can be derived with statistical and systematic uncertainties amounting to better than 0.1\,dex. Where available, a match of abundances from different ionic species is typically achieved.

Phosphorus abundances were derived for a sample of 42 slowly rotating OB-type stars and B-type supergiants based on high-quality spectra, including a subset of 28 stars in the solar neighbourhood, allowing us to constrain the cosmic abundance of phosphorus to $\log$\,(P/H)\,+\,12\,=\,5.36$\pm$0.14. Some first applications of the data were the derivation of the amount of phosphorus depleted onto interstellar dust ($\sim$40\% of the pristine abundance) and the placing of massive star data in the context of the galactochemical evolution of the phosphorus abundances as deduced from analyses of solar-type stars. 

We are confident that the data will be useful for many further applications, one example being reliable determinations of mass loss rates in O-type stars from the \ion{P}{v} resonance doublet at $\lambda\lambda$1118 and 1128\,{\AA} \citep{Fullertonetal06}, which requires knowledge of the abundance of phosphorus. Another example may be the provision of a benchmark value for an important chemical species in the context of astrobiological studies. Investigations of the impact of our non-LTE line-formation calculations for the interpretation of phosphorus-rich chemically peculiar stars will be performed in the future.

\begin{acknowledgements}
We thank the anonymous referee for a constructive report that helped to clarify the paper.
P.A.~acknowledges support of this work by grant of a Ph.D. stipend from the Vice Rectorate for Research of the University of Innsbruck.
Based on observations collected at the European Southern Observatory under ESO programmes
076.C-0431(A), 081.A-9005(A), 088.D-0064(A) and 089.D-0975(A). Based on observations collected at the Centro Astron\'omico Hispano Alem\'an at Calar Alto (CAHA), operated jointly by the Max-Planck Institut f\"ur Astronomie and the Instituto de Astrof\'isica de Andaluc\'ia (CSIC), proposal H2005-2.2-016. This research used the facilities of the Canadian Astronomy Data Centre operated by the National Research Council of Canada with the support of the Canadian Space Agency. 
\end{acknowledgements}

\bibliography{refs}

\begin{appendix}
\section{Assessment of the atomic data quality}\label{appendix:A}
The quality of atomic data used for the construction of model atoms and, therefore, used in the solution of the statistical equilibrium equations is decisive for the quality of the model predictions. Obviously, ab initio data should be preferred over approximations, and we want to comment in the following on the quality of the ab initio data used in the present work.

Typical uncertainties in the radiative data calculated with the methods used for the Opacity Project amount on the whole to $\sim$10\%. The oscillator strengths computed with the MCHF method as used in the final spectrum synthesis are typically of benchmark quality; that is, they should be accurate to a few per cent. To the best of our knowledge, no measurements of branching ratios and lifetimes (which would allow oscillator strengths to be derived) obtained with modern methods are available for the phosphorus ions under consideration here. Uncertainties for the collisional excitation data computed with methods used for the Iron Project \citep{Hummeretal93} typically lie in the range $\sim$20-30\%. Collisional ionisation data are the least accurate in our data set (and in non-LTE studies in general) as only order-of-magnitude approximations are available, but fortunately their rates are unimportant overall.

\begin{figure}
    \centering
    \includegraphics[width=.98\linewidth]{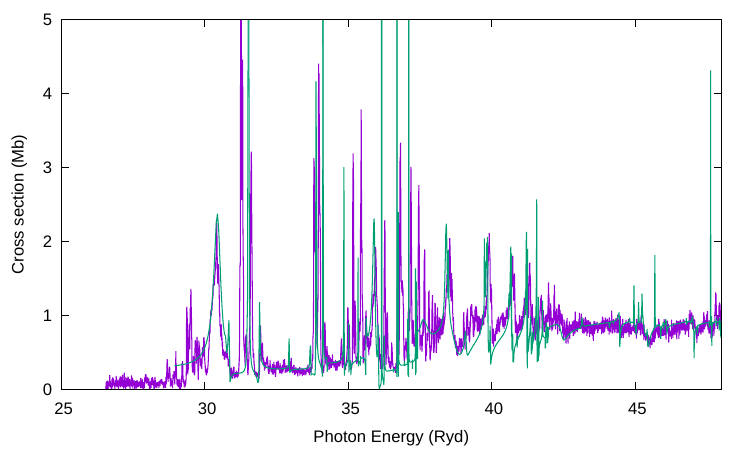}
    \caption{Comparison of the measured \citep[purple line]{HERNANDEZ201580} and calculated \ion{P}{iii} ground state photoionisation cross-section (Butler, in prep., green). The measured data have been scaled by a factor~of~1.3}
    \label{fig:piiicomp}
\end{figure}

\begin{figure}
    \centering
    \includegraphics[width=.98\linewidth]{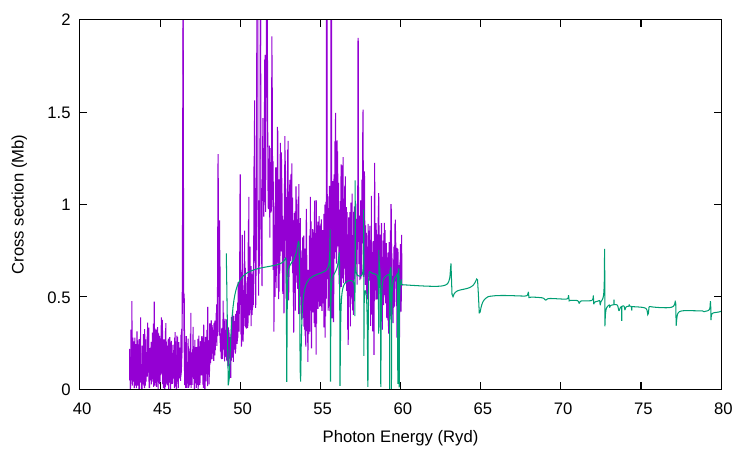}
    \caption{Same as Fig.~\ref{fig:piiicomp}, but for \ion{P}{iv}.}
    \label{fig:pivcomp}
\end{figure}

For the ions investigated here, there exist, however, rarely available measurements: photoionisation cross-sections of the ground states. We have obtained the
data from Nahar's website \citep{Nahar20}. For \ion{P}{iii} \citep{HERNANDEZ201580}, Fig~\ref{fig:piiicomp}, the agreement is excellent while
that for \ion{P}{iv}, Fig~\ref{fig:pivcomp}, is reasonable. Both observations have been scaled by a
factor of 1.3, since the measurements are difficult and the
absolute value, in particular, is subject to error. However, the shape of the curve is relatively well defined.

\begin{figure}
    \centering
    \includegraphics[width=.98\linewidth]{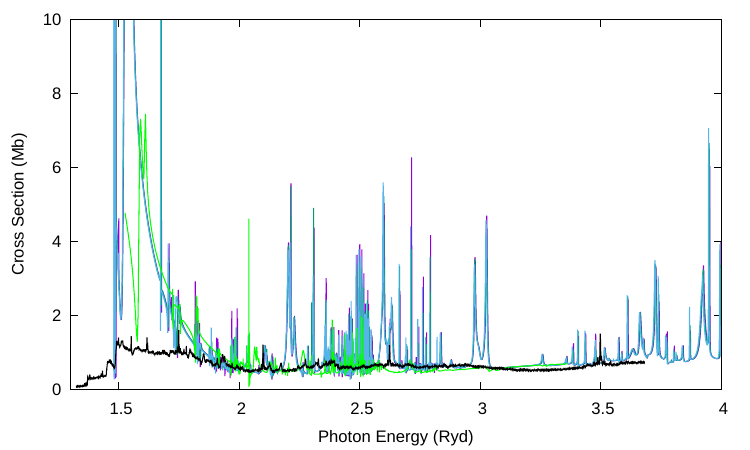}
    \caption{Observed and calculated \citep{NAHAR2017215} \ion{P}{ii}
      ground state photoionisation cross-section. The observed cross-section
      (black) has been scaled by a factor of 0.5. All three cross-sections calculated by Nahar are shown ($J$\,=\,0, 1, 2 in purple, green, blue, respectively)
    together with our own $LS$-coupling calculation (light green).}
    \label{fig:piicomp}
\end{figure}

The ground state cross-section of \ion{P}{ii} has been
determined experimentally by \citet{NAHAR2017215}, with
theoretical values reported in the same work. This uses the
Breit-Pauli approximation and, therefore,
includes some relativistic effects. They are compared with our $LS$-coupling
results in Fig.~\ref{fig:piicomp}. We note that the differences between the three
Breit-Pauli curves are negligible and can only be seen in a few resonances.
This is because the relativstic effects are small. At higher energies all cross-sections
agree well although we have scaled the observation by a factor of 0.5, again
assuming problems with the normalisation of the measured cross-section.
However, both calculations show large resonances and an overall increase closer to threshold while the
experiment does not. This difference is difficult to understand and is not a
topic for this work.

It should be noted that the radiative data in the present work have also been calculated
 including the Breit-Pauli interaction, but these relativistic data have been
 'packed' to reduce the size of the model atom. We use the $LS$-coupling cross-section itself instead of packing as it agrees extremely well with the Breit-Pauli data including the
 large resonances close to the threshold. The large resonances are not expected to
 affect the radiative transfer since they are a) more than 50\,\AA\ wide
 and b) lie only slightly shortward of the Lyman edge where the flux is low.

\begin{figure}
    \centering
    \includegraphics[width=.98\linewidth]{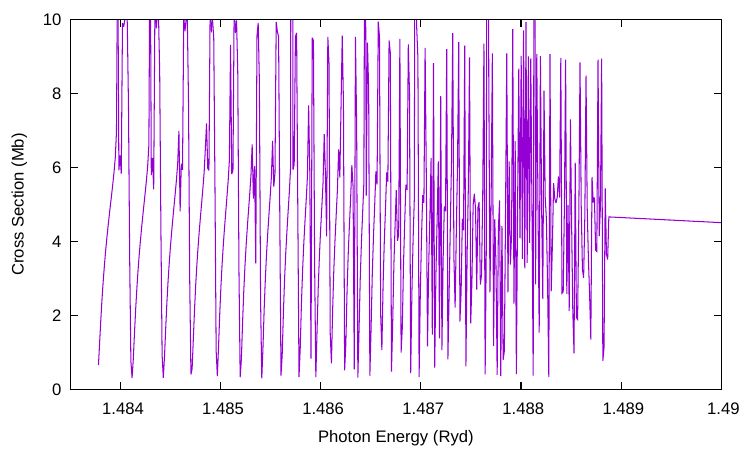}
    \caption{\ion{P}{ii} ground state photoionisation cross-section of \citet{NAHAR2017215} just above
      threshold. The regular pattern disappears to higher energies due to the
      energy mesh and replicates the merging of the lines to form an average
      converging to the continuous cross-section above the first excited state above an energy of about 1.489\,Ryd.}
    \label{fig:nearthresh}
\end{figure}

 There is one more relativistic effect which should be mentioned. In the
 region just above \ion{P}{ii} ground state threshold (see Fig.~\ref{fig:nearthresh}) 
 there is a series of
 resonances converging to the first excited level of \ion{P}{iii}. These do
 not appear in the $LS$-coupling version. However, the lowest of these
 resonances has a principal quantum number of roughly 28 so that it and the higher values
 will be pressure broadened and form a quasi-continuum with an average similar
 to that given by the $LS$-coupling value, justifying the use of the latter. We mention this because a straightforward use of the Breit-Pauli cross-section of \citet{NAHAR2017215} can affect the radiative transfer. The photoionisation rate will be sensitive to the details of the sampling of the cross-section. That is, care has to be taken in establishing the frequency grid in order to avoid maxima or minima of the resonances. Otherwise the rate can become inaccurate.

We emphasise that these remarks apply to all light ions in which the fine
 structure splittings are small, which is in any case a requirement if the
 energy levels are to be packed.

 Overall, the uncertainties of the ab initio data are small enough to not affect the abundance analysis as shown in previous work \citep{Przybillaetal00,Przybillaetal01a,Przybillaetal01b,PrBu01}.

\section{Summary of analysed stars}\label{appendix:B}
In Table~\ref{table:results} we summarise all the stars analysed in this work. For each object its spectral type is given,
the instrument(s) used for the observations, its relevant atmospheric parameters (effective temperature, surface gravity,
microturbulent, projected rotational, and macroturbulent velocities), the reference for the atmospheric parameters,
the distance to the object and the non-LTE phosphorus abundance. The table is subdivided into three sections,
concentrating on the late O-type main-sequence stars among the sample objects,
the early B-type main-sequence stars and the evolved B-type supergiants.

\begin{table*}
\centering
\caption{Properties of the sample stars and non-LTE phosphorus abundances.}
\label{table:results}
\small
\setlength{\tabcolsep}{1.5mm}
\begin{tabular}{rlllllrrrcll}
\hline\hline
\# & Object & Sp. Type & Instrument & $T_\mathrm{eff}$ & $\log g$ & \multicolumn{1}{c}{$\xi$} & $\varv \sin i$ & \multicolumn{1}{c}{$\zeta$} & Ref. & $d$\tablefootmark{a} & $\log \mathrm{P/H}$\,+\,12 \\
\cline{7-9}
 & & & & K & (cgs) &\multicolumn{3}{c}{km\,s$^{-1}$} & & pc \\
\hline
\multicolumn{5}{l}{late O-type stars:}\\
 & HD\,34078     & O9.5\,V    & FOCES/UVES & 33200$\pm$300 & 4.06$\pm$0.05 &  8$\pm$2 &  9$\pm$2  & 23$\pm$3  & 1 & 380$^{+5}_{-5}$ & 5.40$\pm$0.03\,(3) \\
 & HD\,36512     & O9.7\,V    & FOCES/UVES & 32900$\pm$600 & 4.20$\pm$0.05 &  7$\pm$2 &  8$\pm$4  & 26$\pm$4  & 1 & 408$^{+23}_{-27}$ & 5.35$\pm$0.05\,(3) \\
 & HD\,46202     & O9.2\,V    & FEROS/UVES & 33900$\pm$500 & 4.16$\pm$0.05 &  6$\pm$2 & 11$\pm$1  & 33$\pm$4  & 1 & 1442$^{+685}_{-180}$ & 5.30$\pm$0.03\,(3) \\
 & HD\,54879     & O9.7\,V    & FEROS      & 32200$\pm$700 & 4.06$\pm$0.05 &  4$\pm$2 &  0$\pm$1  &  0$\pm$1  & 1 & 1198$^{+45}_{-41}$ & 5.27\,(1) \\
 & HD\,57682     & O9.2\,IV   & UVES       & 33500$\pm$800 & 3.93$\pm$0.05 &  4$\pm$2 & 10$\pm$1  & 33$\pm$4  & 1 & 1106$^{+111}_{-82}$ & 5.30$\pm$0.07\,(3) \\
 & HD\,206183    & O9.5\,V    & FOCES      & 33100$\pm$300 & 4.06$\pm$0.05 &  4$\pm$2 &  4$\pm$2  & 21$\pm$4  & 1 & 903$^{+14}_{-15}$ & 5.18\,(1) \\
 & HD\,207538    & O9.7\,IV   & ESPaDOnS   & 31400$\pm$500 & 3.93$\pm$0.05 &  8$\pm$2 & 32$\pm$3  & 42$\pm$3  & 1 & 834$^{+11}_{-10}$ & 5.36\,(1) \\
 & HD\,214680    & O9\,V      & FOCES      & 34550$\pm$300 & 4.04$\pm$0.05 &  5$\pm$2 & 14$\pm$1  & 32$\pm$2  & 1 & 454$^{+25}_{-32}$ & 5.25\,(1) \\
 & BD\,-13\,4930 & O9.5\,V    & ESPaDOnS   & 32900$\pm$300 & 4.14$\pm$0.05 &  6$\pm$2 &  8$\pm$2  & 23$\pm$4  & 1 & 1579$^{+45}_{-46}$ & 5.41\,(1) \\
\multicolumn{5}{l}{early B-type stars:}\\
 & HD\,886       & B2\,IV     & FOCES      & 22000$\pm$400 & 3.95$\pm$0.05 &  2$\pm$1 &  9$\pm$2  &  8$\pm$2  & 2 & 146$^{+10}_{-8}$ & 5.25$\pm$0.02\,(3) \\
 & HD\,3360      & B2\,IV     & FIES       & 20750$\pm$200 & 3.80$\pm$0.05 &  2$\pm$1 & 20$\pm$2  & 12$\pm$5  & 2 & 150$^{+7}_{-8}$ & 5.42$\pm$0.07\,(3) \\
 & HD\,16582     & B2\,IV     & CAFE       & 21250$\pm$400 & 3.80$\pm$0.05 &  2$\pm$1 & 15$\pm$2  & 10$\pm$5  & 2 & 194$^{+10}_{-9}$ & 5.45$\pm$0.10\,(2) \\
 & HD\,34816     & B0.5\,IV   & FEROS      & 30000$\pm$300 & 4.15$\pm$0.08 &  4$\pm$1 & 25$\pm$2  & 32$\pm$7  & 2\tablefootmark{b} & 298$^{+19}_{-14}$ & 5.67$\pm$0.05\,(3) \\
 & HD\,35039     & B2\,V      & FIES       & 19600$\pm$200 & 3.56$\pm$0.07 &  4$\pm$1 & 12$\pm$1  &  7$\pm$1  & 3 & 350$^{+22}_{-16}$ & 5.31$\pm$0.06\,(5) \\
 & HD\,35299     & B1.5\,V    & FOCES      & 23500$\pm$300 & 4.20$\pm$0.05 &  0$\pm$1 &  8$\pm$1  &    ...    & 2 & 356$^{+12}_{-11}$ & 5.18$\pm$0.06\,(3) \\
 & HD\,35708     & B2.5\,IV   & FOCES      & 20700$\pm$200 & 4.15$\pm$0.07 &  2$\pm$1 & 25$\pm$2  & 17$\pm$5  & 2 & 185$^{+5}_{-4}$ & 5.46$\pm$0.01\,(3) \\
 & HD\,35912     & B2\,V      & FIES       & 19800$\pm$400 & 4.20$\pm$0.10 &  0$\pm$1 &  8$\pm$4  & 18$\pm$5  & 3\tablefootmark{b} & 365$^{+10}_{-9}$ & 5.30$\pm$0.02\,(2) \\
 & HD\,36285     & B2\,V      & FIES       & 21700$\pm$300 & 4.25$\pm$0.08 &  0$\pm$1 & 11$\pm$1  &  8$\pm$1  & 3 & 351$^{+7}_{-7}$ & 5.39$\pm$0.09\,(3) \\
 & HD\,36591     & B1\,V      & FEROS      & 27000$\pm$300 & 4.12$\pm$0.05 &  3$\pm$1 & 12$\pm$1  &    ...    & 2 & 425$^{+36}_{-31}$ & 5.25$\pm$0.03\,(3) \\
 & HD\,36629     & B2\,V      & FIES       & 20300$\pm$400 & 4.15$\pm$0.10 &  2$\pm$1 & 10$\pm$1  &  5$\pm$1  & 3 & 424$^{+8}_{-6}$ & 5.30$\pm$0.03\,(3) \\
 & HD\,36822     & B0\,III    & FOCES/UVES & 30000$\pm$300 & 4.05$\pm$0.10 &  8$\pm$1 & 28$\pm$2  & 18$\pm$5  & 2 & 374$^{+45}_{-36}$ & 5.48$\pm$0.08\,(3) \\
 & HD\,36959     & B1\,V      & FIES       & 26100$\pm$200 & 4.25$\pm$0.07 &  0$\pm$1 & 12$\pm$1  &  5$\pm$1  & 3 & 358$^{+15}_{-14}$ & 5.24$\pm$0.03\,(2) \\
 & HD\,37042     & B0.7\,V    & FIES       & 29300$\pm$300 & 4.30$\pm$0.09 &  2$\pm$1 & 30$\pm$2  & 10$\pm$3  & 3 & 413$^{+11}_{-8}$ & 5.52\,(1) \\
 & HD\,37744     & B1.5\,V    & FIES       & 24600$\pm$300 & 4.20$\pm$0.05 &  0$\pm$1 & 37$\pm$1  &  3$\pm$2  & 3\tablefootmark{b} & 389$^{+10}_{-12}$ & 5.13$\pm$0.02\,(2) \\
 & HD\,44743     & B1\,II/III & FEROS      & 24300$\pm$500 & 3.60$\pm$0.05 &  8$\pm$1 & 30$\pm$2  & 32$\pm$3  & 4\tablefootmark{b} & 151$^{+5}_{-5}$ & 5.12$\pm$0.04\,(2) \\
 & HD\,63922     & B0\,III    & FEROS      & 31200$\pm$300 & 3.95$\pm$0.05 &  8$\pm$1 & 29$\pm$4  & 37$\pm$8  & 2 & 297$^{+52}_{-31}$ & 5.37\,(1) \\
 & HD\,122980    & B2\,V      & FEROS      & 20800$\pm$300 & 4.22$\pm$0.05 &  3$\pm$1 & 18$\pm$1  &    ...    & 2 & 153$^{+5}_{-5}$ & 5.21$\pm$0.03\,(3) \\
 & HD\,149438    & B0.2\,V    & FEROS/UVES & 32000$\pm$300 & 4.30$\pm$0.05 &  5$\pm$1 &  4$\pm$1  &  4$\pm$1  & 2 & 145$^{+12}_{-10}$ & 5.38$\pm$0.08\,(4) \\
 & HD\,160762    & B3\,IV     & FIES       & 17500$\pm$200 & 3.80$\pm$0.05 &  1$\pm$1 &  6$\pm$1  &    ...    & 2 & 153$^{+6}_{-4}$ & 5.40$\pm$0.13\,(9) \\
 & HD\,205021    & B1\,IV     & FOCES      & 27000$\pm$450 & 3.95$\pm$0.05 &  4$\pm$1 & 28$\pm$3  & 20$\pm$7  & 2 & 210$^{+14}_{-12}$ & 5.41\,(1) \\
 & HD\,209008    & B3\,III    & FOCES      & 15800$\pm$200 & 3.75$\pm$0.05 &  4$\pm$1 & 15$\pm$3  & 10$\pm$3  & 2 & 459$^{+22}_{-19}$ & 5.56$\pm$0.05\,(4) \\
 & HD\,216916    & B2\,IV     & FOCES      & 23000$\pm$200 & 3.95$\pm$0.05 &  0$\pm$1 & 12$\pm$1  &    ...    & 2 & 441$^{+18}_{-14}$ & 5.26\,(1) \\
\multicolumn{5}{l}{B-type supergiants:}\\
 & HD\,14818     & B2\,Ia     & FOCES      & 18600$\pm$300 & 2.45$\pm$0.07 & 14$\pm$2 & 48$\pm$6  & 40$\pm$5  & 6 & 2121$^{+145}_{-107}$ & 5.16\,(1) \\
 & HD\,34085     & B8\,Ia     & FEROS      & 12000$\pm$200 & 1.75$\pm$0.10 &  7$\pm$1 & 36$\pm$5  & 22$\pm$5  & 5 & 265$^{+26}_{-22}$ & 5.53$\pm$0.07\,(6) \\
 & HD\,51309     & B3\,Ib     & FEROS      & 15600$\pm$400 & 2.59$\pm$0.05 & 10$\pm$2 & 30$\pm$6  & 35$\pm$5  & 6 & 1108$^{+407}_{-215}$ & 5.30$\pm$0.23\,(4) \\
 & HD\,119646    & B1.5\,Ib   & FEROS      & 19700$\pm$200 & 2.90$\pm$0.07 & 14$\pm$2 & 37$\pm$6  & 40$\pm$5  & 6 & 1721$^{+84}_{-73}$ & 5.06\,(1) \\
 & HD\,125288    & B5\,II     & FEROS      & 13700$\pm$300 & 2.77$\pm$0.05 &  6$\pm$2 & 23$\pm$4  & 30$\pm$5  & 6 & 438$^{+51}_{-30}$ & 5.63$\pm$0.04\,(3) \\
 & HD\,159110    & B2\,II     & FEROS      & 19500$\pm$300 & 3.42$\pm$0.05 &  3$\pm$2 & 17$\pm$4  & 15$\pm$5  & 6 & 1362$^{+80}_{-71}$ & 5.27$\pm$0.02\,(3) \\
 & HD\,164353    & B5\,Ib     & FEROS      & 14700$\pm$300 & 2.57$\pm$0.05 &  8$\pm$2 & 20$\pm$4  & 32$\pm$5  & 6 & 797$^{+196}_{-129}$ & 5.33$\pm$0.06\,(3) \\
 & HD\,183143    & B7\,Ia     & FOCES      & 12800$\pm$200 & 1.76$\pm$0.05 &  7$\pm$2 & 37$\pm$5  & 27$\pm$5  & 6 & 2168$^{+119}_{-122}$ & 5.59$\pm$0.05\,(2) \\
 & HD\,184943    & B8\,Iab    & FOCES      & 11900$\pm$200 & 1.88$\pm$0.05 &  9$\pm$2 & 35$\pm$6  & 25$\pm$5  & 6 & 4090$^{+236}_{-236}$ & 5.52\,(1) \\
 & HD\,191243    & B5\,Ib     & FOCES      & 14000$\pm$300 & 2.64$\pm$0.06 &  8$\pm$2 & 27$\pm$6  & 25$\pm$5  & 6 & 1205$^{+32}_{-32}$ & 5.56$\pm$0.05\,(4) \\
\hline
\end{tabular}
\tablebib{(1) \cite{Aschenbrenneretal23}; (2) \cite{NiPr12}; (3) \cite{NiSi11}; (4) \cite{Fossatietal15}; (5) \cite{Przybillaetal2006}; (6) \cite{Wessmayeretal22}.}
\tablefoot{The number in brackets at the end of each row gives the number of phosphorus lines analysed.
\tablefoottext{a}{\cite{Gaia2016, Gaia21} photogeometric distances and associated 14th and 86th confidence percentiles \citep{BailerJonesetal21}; for HD\,36591, HD\,36822 and HD\,36959 inverse parallax distances based on \cite{Gaia2023} data; for HD\,149438, HD\,205021, HD\,44743 and HD\,34085 inverse parallax distances based on \cite{vanLeeuwen07a} data.}
\tablefoottext{b}{The atmospheric parameters of these stars were refined in this work. If available, the iron abundances to be used in Fig.~\ref{fig:P_FE} were adopted from the original references.}
}
\end{table*}

\section{Line-by-line phosphorus abundances}\label{appendix:C}
Table~\ref{table:line_fits} provides information on the non-LTE abundances derived from all individual phosphorus lines analysed, for all sample stars. The stars are ordered using the same scheme as established in Table~\ref{table:results}.

\clearpage

\setcounter{section}{3}

\begin{sidewaystable*}
\caption{Line-by-line phosphorus abundance for each analysed star.}
\label{table:line_fits}
\centering
\resizebox{\textwidth}{!}{
\begin{tabular}{lcccccccccccccccc} 
\hline\hline
       & \multicolumn{7}{c}{\ion{P}{II}} & & \multicolumn{3}{c}{\ion{P}{III}} & & \multicolumn{4}{c}{\ion{P}{IV}} \\
\cline{2-8} \cline{10-12} \cline{14-17}
Object & 4499.230 & 5253.479 & 5296.077 & 5386.895 & 5425.880 & 6024.178 & 6043.084 & & 4059.312 & 4080.084 & 4246.720 & & 3347.739 & 3364.470 & 3371.119 & 4249.655 \\
\hline
  HD\,34078     & ...  & ...  & ...  & ...  & ...  & ...  & ...  & & ...  & ...  & ...  & & 5.41 & 5.42 & ...  & 5.36  \\
  HD\,36512     & ...  & ...  & ...  & ...  & ...  & ...  & ...  & & ...  & ...  & ...  & & 5.30 & 5.36 & ...  & 5.39  \\
  HD\,46202     & ...  & ...  & ...  & ...  & ...  & ...  & ...  & & ...  & ...  & ...  & & 5.33 & 5.31 & ...  & 5.27  \\
  HD\,54879     & ...  & ...  & ...  & ...  & ...  & ...  & ...  & & ...  & ...  & ...  & & ...  & ...  & ...  & 5.27  \\
  HD\,57682     & ...  & ...  & ...  & ...  & ...  & ...  & ...  & & ...  & ...  & ...  & & 5.30 & 5.37 & ...  & 5.24  \\
  HD\,206183    & ...  & ...  & ...  & ...  & ...  & ...  & ...  & & ...  & ...  & ...  & & ...  & ...  & ...  & 5.18  \\
  HD\,207538    & ...  & ...  & ...  & ...  & ...  & ...  & ...  & & ...  & ...  & ...  & & ...  & ...  & ...  & 5.36  \\
  HD\,214680    & ...  & ...  & ...  & ...  & ...  & ...  & ...  & & ...  & ...  & ...  & & ...  & ...  & ...  & 5.25  \\
  BD\,-13\,4930 & ...  & ...  & ...  & ...  & ...  & ...  & ...  & & ...  & ...  & ...  & & ...  & ...  & ...  & 5.41  \\[2mm]
  HD\,886       & ...  & ...  & ...  & ...  & ...  & ...  & ...  & & 5.27 & 5.23 & 5.24 & & ...  & ...  & ...  & ...   \\
  HD\,3360      & ...  & ...  & ...  & ...  & ...  & ...  & ...  & & 5.35 & 5.42 & 5.49 & & ...  & ...  & ...  & ...   \\
  HD\,16582     & ...  & ...  & ...  & ...  & ...  & ...  & ...  & & 5.38 & ...  & 5.52 & & ...  & ...  & ...  & ...   \\
  HD\,34816     & ...  & ...  & ...  & ...  & ...  & ...  & ...  & & 5.67 & ...  & 5.72 & & ...  & ...  & ...  & 5.62  \\
  HD\,35039     & ...  & 5.29 & ...  & ...  & 5.40 & ...  & ...  & & 5.23 & 5.31 & 5.31 & & ...  & ...  & ...  & ...   \\
  HD\,35299     & ...  & ...  & ...  & ...  & ...  & ...  & ...  & & 5.12 & 5.19 & 5.23 & & ...  & ...  & ...  & ...   \\
  HD\,35708     & ...  & ...  & ...  & ...  & ...  & ...  & ...  & & 5.45 & 5.47 & 5.45 & & ...  & ...  & ...  & ...   \\
  HD\,35912     & ...  & ...  & ...  & ...  & ...  & ...  & ...  & & 5.29 & ...  & 5.32 & & ...  & ...  & ...  & ...   \\
  HD\,36285     & ...  & ...  & ...  & ...  & ...  & ...  & ...  & & 5.49 & 5.38 & 5.31 & & ...  & ...  & ...  & ...   \\
  HD\,36591     & ...  & ...  & ...  & ...  & ...  & ...  & ...  & & 5.28 & 5.23 & 5.23 & & ...  & ...  & ...  & ...   \\
  HD\,36629     & ...  & ...  & ...  & ...  & ...  & ...  & ...  & & 5.30 & 5.33 & 5.28 & & ...  & ...  & ...  & ...   \\
  HD\,36822     & ...  & ...  & ...  & ...  & ...  & ...  & ...  & & ...  & ...  & ...  & & 5.41 & 5.47 & ...  & 5.57  \\
  HD\,36959     & ...  & ...  & ...  & ...  & ...  & ...  & ...  & & 5.26 & ...  & 5.22 & & ...  & ...  & ...  & ...   \\
  HD\,37042     & ...  & ...  & ...  & ...  & ...  & ...  & ...  & & 5.52 & ...  & ...  & & ...  & ...  & ...  & ...   \\
  HD\,37744     & ...  & ...  & ...  & ...  & ...  & ...  & ...  & & 5.15 & ...  & 5.12 & & ...  & ...  & ...  & ...   \\
  HD\,44743     & ...  & ...  & ...  & ...  & ...  & ...  & ...  & & 5.09 & ...  & 5.14 & & ...  & ...  & ...  & ...   \\
  HD\,63922     & ...  & ...  & ...  & ...  & ...  & ...  & ...  & & ...  & ...  & ...  & & ...  & ...  & ...  & 5.37  \\
  HD\,122980    & ...  & ...  & ...  & ...  & ...  & ...  & ...  & & 5.24 & 5.19 & 5.20 & & ...  & ...  & ...  & ...   \\
  HD\,149438    & ...  & ...  & ...  & ...  & ...  & ...  & ...  & & ...  & ...  & ...  & & 5.33 & 5.29 & 5.43 & 5.47  \\
  HD\,160762    & 5.59 & 5.24 & ...  & 5.56 & 5.30 & 5.42 & 5.46 & & 5.22 & 5.46 & 5.39 & & ...  & ...  & ...  & ...   \\
  HD\,205021    & ...  & ...  & ...  & ...  & ...  & ...  & ...  & & 5.41 & ...  & ...  & & ...  & ...  & ...  & ...   \\
  HD\,209008    & ...  & 5.58 & 5.61 & ...  & 5.49 & ...  & 5.56 & & ...  & ...  & ...  & & ...  & ...  & ...  & ...   \\
  HD\,216916    & ...  & ...  & ...  & ...  & ...  & ...  & ...  & & ...  & ...  & 5.26 & & ...  & ...  & ...  & ...   \\[2mm]
  HD\,14818     & ...  & ...  & ...  & ...  & ...  & ...  & ...  & & ...  & ...  & 5.16 & & ...  & ...  & ...  & ...   \\
  HD\,34085     & ...  & 5.54 & 5.56 & ...  & 5.55 & ...  & 5.46 & & 5.43 & ...  & 5.61 & & ...  & ...  & ...  & ...   \\
  HD\,51309     & ...  & ...  & ...  & ...  & 5.51 & ...  & 5.49 & & 5.05 & ...  & 5.15 & & ...  & ...  & ...  & ...   \\
  HD\,119646    & ...  & ...  & ...  & ...  & ...  & ...  & ...  & & ...  & ...  & 5.06 & & ...  & ...  & ...  & ...   \\
  HD\,125288    & ...  & ...  & 5.58 & 5.65 & ...  & ...  & 5.66 & & ...  & ...  & ...  & & ...  & ...  & ...  & ...   \\
  HD\,159110    & ...  & ...  & ...  & ...  & ...  & ...  & ...  & & 5.25 & 5.26 & 5.29 & & ...  & ...  & ...  & ...   \\
  HD\,164353    & ...  & ...  & ...  & ...  & ...  & ...  & ...  & & 5.32 & 5.39 & 5.28 & & ...  & ...  & ...  & ...   \\
  HD\,183143    & ...  & ...  & ...  & ...  & 5.56 & ...  & 5.63 & & ...  & ...  & ...  & & ...  & ...  & ...  & ...   \\
  HD\,184943    & ...  & ...  & ...  & ...  & ...  & ...  & 5.52 & & ...  & ...  & ...  & & ...  & ...  & ...  & ...   \\
  HD\,191243    & ...  & 5.61 & 5.60 & ...  & 5.55 & ...  & ...  & & ...  & ...  & 5.49 & & ...  & ...  & ...  & ...   \\
\hline
\end{tabular}
}
\end{sidewaystable*}
\end{appendix}

\end{document}